\documentclass[pra,aps,10pt]{revtex4-2}
\usepackage{amsmath,amssymb,graphicx}
\usepackage{color}
\usepackage{subfigure}

\begin{document}

\title{Nonlinear dynamics of wave packets in tunnel-coupled
harmonic-oscillator traps}
\author{Nir Hacker$^1$ and Boris A. Malomed$^{1,2}$}
\affiliation{$^1$Department of Physical Electronics, School of Electrical Engineering, Faculty of
Engineering, and Center for Light-Matter interaction, Tel Aviv University,
Tel Aviv 69978, Israel}
\affiliation{$^2$Instituto de Alta
Investigaci\'{o}n, Universidad de Tarapac\'{a}, Casilla 7D, Arica, Chile}

\begin{abstract}
We consider a two-component linearly-coupled system with the intrinsic cubic
nonlinearity and the harmonic-oscillator (HO) confining potential. The
system models binary settings in BEC\ and optics. In the symmetric system,
with the HO trap acting in both components, we consider Josephson
oscillations (JO) initiated by an input in the form of the HO's ground state
(GS) or dipole mode (DM), placed in one component. With the increase of the
strength of the self-focusing nonlinearity, spontaneous symmetry breaking
(SSB) between the components takes place in the dynamical JO state. Under
still stronger nonlinearity, the regular JO initiated by the GS input carry
over into a chaotic dynamical state. For the DM input, the chaotization
happens at smaller powers than for the GS, which is followed by SSB at a
slightly stronger nonlinearity. In the system with the defocusing
nonlinearity, SSB does not take place, and dynamical chaos occurs in a small
area of the parameter space. In the asymmetric half-trapped system, with the
HO potential applied to a single component, we first focus on the spectrum
of confined binary modes in the linearized system. The spectrum is found
analytically in the limits of weak and strong inter-component coupling, and
numerically in the general case. Under the action of the coupling, the
existence region of the confined modes shrinks for GSs and expands for DMs.
In the full nonlinear system, the existence region for confined modes is
identified in the numerical form. They are constructed too by means of the
Thomas-Fermi approximation, in the case of the defocusing nonlinearity.
Lastly, particular (non-generic) exact analytical solutions for confined
modes, including vortices, in one- and two-dimensional asymmetric linearized
systems are found. They represent bound states in the continuum.
\end{abstract}

\maketitle

\section{Introduction}

The combination of a harmonic-oscillator (HO) trapping potential and cubic
nonlinearity is a ubiquitous setting which occurs in diverse microscopic,
mesoscopic, and macroscopic physical settings. A well-known realization is
offered by Bose-Einstein condensates (BECs) with collisional nonlinearity
\cite{Pet,Pit,Pan}, loaded in a magnetic or optical trap -- see, e.g., Refs.~\cite{Schneider}-\cite{Newcastle}. A similar combination of the effective
confinement, approximated by the parabolic profile of the local refractive
index, and the Kerr term is relevant as a model of optical waveguides \cite%
{Agrawal}-\cite{Thaw}. Models of the same type appear in other physical
systems too, such as networks of Josephson oscillators \cite{Josephson
network}.

The character of states created by the interplay of the intrinsic
nonlinearity and externally applied trapping potential strongly depends on
the sign of the nonlinearity. In the case of the self-attraction (or
self-phase-modulation, SPM, in terms of optics \cite{KA}), localized modes,
similar to solitons, arise spontaneously. On the other hand, self-repulsion
tends to create flattened configurations, which, in turn, may support dark
solitons in various static and dynamical states \cite{Busch}-\cite{Newcastle}%
. A specific situation occurs in a system combining repulsion and a weak
parabolic potential with an additional spatially periodic one (it represents
an optical lattice in BEC \cite{Morsch}, or a photonic crystal in optics and
plasmonics \cite{PhCr1,PhCr2,PhCr3}): the interplay of the periodic
potential with the self-repulsion gives rise to bright gap solitons \cite%
{KB,Morsch}, whose effective mass is negative. For this reason, gap solitons
are expelled by the HO potential, but are trapped by the inverted one, that
would expel modes with positive masses \cite{HS}.

Another noteworthy feature of the dynamics of one-dimensional (1D) nonlinear
fields trapped in confining potentials is the degree of its \textit{%
nonintegrability}. The generic model for such settings is provided by the
nonlinear Schr\"{o}dinger equation (NLSE, alias the Gross-Pitaevskii
equation, in terms of BEC \cite{Pet,Pit,Pan}) with the cubic term, whose
integrability in the 1D space \cite{Zakharov} is broken by the presence of
the HO potential. Nevertheless, systematic numerical simulations, performed
for the repulsive sign of the cubic nonlinearity, have demonstrated that the
long-time evolution governed by NLSE with the HO potential term does not
leads to establishment of spatiotemporal chaos (``turbulence"), which would
be expected in the case of generic nonintegrability \cite{turbulence,Mazets}%
. Instead, the setup demonstrates quasi-periodic evolution, represented by a
quasi-discrete power spectrum, in terms of a multi-mode truncation (Galerkin
approximation) \cite{Newcastle}. This observation is specific for the
harmonic (quadratic)\ confining potential, while anharmonic ones quickly
lead to the onset of clearly observed chaos \cite{Nick1,Mazets2}.
Thermalization of the model with the HO potential was recently explored, in
the framework of a stochastically driven dissipative Gross-Pitaevskii
equation, in Ref.~\cite{Kheruntsyan}.

The interplay of the cubic nonlinearity and trapping potentials also occurs
in two-component systems, which represent, in particular, binary BEC \cite%
{Brazhnyi,Merhasin,bright-dark,Viskol}. Here, we aim to consider the system
with linear coupling between the components. In optics, if two modes
carrying orthogonal polarizations of light propagate in the same waveguide,
linear mixing between them is induced by a twist of the guiding structure,
see, e.g., Ref.~\cite{twist}. In BEC, different atomic states which
correspond to the interacting components can be linearly coupled by a
resonant electromagnetic wave \cite{GHz1}-\cite{GHz4}. Another realization
of linearly-mixed systems is offered by dual-core waveguides, coupled by
tunneling of the field across a barrier separating the cores. The dual-core
schemes are equally relevant to optics and BEC \cite{book}. In particular,
experiments with temporal solitons in dual-core optical fibers were reported
in recent work \cite{Ignac}.

The coupling between the components enhances the complexity of the system
and makes it possible to find new static and dynamical states in it. In
particular, the symmetric system combining the attractive cubic terms of the
SPM type and (optionally) HO potential acting in each component, with linear
coupling between them, gives rise to spontaneous symmetry breaking (SSB) of
two-component states \cite{Viskol,Ignac}. In the case of repulsive SPM
acting in each component and nonlinear repulsion between them (cross-phase
modulation, XPM), it was found \cite{Merhasin} that the linear mixing shifts
the miscibility-immiscibility transition \cite{Mineev} in the trapped
condensate. Further, effects of nonintegrability may be stronger in the
linearly coupled system, because the linear coupling makes the system of
one-dimensional NLSEs nonintegrable, even in the absence of the HO potential
\cite{OptJO6}, although the integrability is kept by the system with the
linear coupling added to the SPM and XPM terms with equal strengths (the
Manakov's nonlinearity) \cite{ST}.

The objective of the present work is two-fold. First, in Section II we aim
to analyze the onset of chaos, as well as SSB, in the symmetric
linearly-coupled system, with both attractive and repulsive signs of the SPM
terms, starting from an input in the form of the ground state (GS), or the
first excited state (the dipole mode (DM), represented by a spatially odd
wave function) of the HO, which is initially launched in one core
(component), while the other one is empty. This type of the input is, in
particular, experimentally relevant in optics \cite{Ignac}. As a nonchaotic
dynamical regime in this case, one may expect Josephson oscillations (JO) of
the optical field \cite{OptJO1}-\cite{OptJO4}, \cite{OptJO6}, \cite{Ignac},
or of the BEC wave function \cite{BECJO1}-\cite{HS2}, between the cores. As
a measure of the transition to dynamical chaos in the system, we use a
relative spread of the power spectrum of oscillations produced by
simulations of the coupled NLSEs. Naturally, the chaotization sets in above
a certain threshold value of the nonlinearity strength, and the chaos is
much weaker in the case of the repulsive nonlinearity. In the dynamical
state initiated by the GS, the SSB takes place prior to the onset of the
dynamical chaos, while the DM input undergoes the chaotization occurs first,
followed by the SSB at a slightly stronger nonlinearity.

The second objective, which is presented below in Section III, is to
construct stationary GS and DM in the asymmetric (\textit{half-trapped})
linearly-coupled system, with the HO potential applied to one component
only. The latter system can be realized in the experiment, applying, for
instance, the trapping potential only to one core of the double waveguide
for matter waves in BEC (e.g., by focusing laser beams, which induced the
trapping, on the single core). In optics, a similar setup may be built as a
coupler with two widely different cores, narrow and broad ones, with the
narrow core emulating the component carrying a tightly confining potential,
cf. Ref.~\cite{Osgood}. A remarkable peculiarity of such a system is that
the linear coupling mixes completely different types of the asymptotic
behavior at $|x|\rightarrow \infty $: the trapped component is always
confined, in the form of a Gaussian, by the HO potential, while the
untrapped one is free to escape. In addition, the asymmetry between the
coupled cores makes it necessary to take into regard a difference in the
chemical potentials or propagation constants (in terms of the BEC and
optics, respectively) between them, which is represented by parameters $%
\omega $ in Eq.~(\ref{asymm}), see below. Some results for the half-trapped
system are obtained in an analytical form, in the weak- and strong-coupling
limits, as well as by means of the Thomas-Fermi approximation (TFA), and
full results are produced numerically. In addition, particular solutions for
localized states of the linear half-trapped system (including vortex states
in its 2D version) are found in an exact form.\ The exact solutions belong
to the class of \emph{bound states in the continuum} \cite{BIC,BIC2,BIC3},
alias \emph{embedded} \cite{embedded} ones, which are spatially confined
modes existing, as exceptional states, with the carrier frequency falling in
the continuous spectrum. The paper is concluded by Section IV.

\section{The symmetric system}

\subsection{The coupled equations}

As outlined above, we consider systems modeled by a pair of coupled NLSEs
for complex amplitudes $u\left( x,z\right) $ and $v\left( x,z\right) $ of
two interacting waves. In the normalized form, the equations are written in
terms of the spatial-domain propagation in an optical waveguide with
propagation distance $z$ and transverse coordinate $x$:%
\begin{eqnarray}
i\frac{\partial u}{\partial z}+\frac{1}{2}\frac{\partial ^{2}u}{\partial
x^{2}}+\lambda v-\frac{\Omega ^{2}}{2}x^{2}u+\sigma \left(
|u|^{2}+g|v|^{2}\right) u &=&0,  \notag \\
&&  \label{uv} \\
i\frac{\partial v}{\partial z}+\frac{1}{2}\frac{\partial ^{2}v}{\partial
x^{2}}+\lambda u-\frac{\Omega ^{2}}{2}x^{2}v+\sigma \left(
|v|^{2}+g|u|^{2}\right) v &=&0.  \notag
\end{eqnarray}%
Here, coefficients $\sigma >0$ or $\sigma <0$ represents the strength of the
focusing or defocusing SPM in each core, while $\lambda $ and $\sigma g$
represent the linear mixing and XPM interaction, respectively. By means of
rescaling, the strength of the HO trapping potential is set to be $\Omega =1$
(unless it is zero in one core). In other words, $x$ is measured in units of
the respective HO length. This implies that the unit of the transverse
coordinate in the optical waveguides takes typical values in the range of $%
10-30$ $\mathrm{\mu }$m, hence the respective unit of the propagation
distance (the Rayleigh/diffraction distance corresponding to the OH length)
is estimated to be between $1$ mm and $1$ cm, for the carrier wavelength $%
\sim 1$ $\mathrm{\mu }$m. In the matter-wave realization of the system,
typical units of $x$ and time (replacing $z$ in Eq.~(\ref{uv}) are $\sim 10$
$\mathrm{\mu }$m and $10$ ms, respectively.

The system conserves two dynamical invariants., \textit{viz}., the total
norm (or power, in terms of optics),%
\begin{equation}
P=\int_{-\infty }^{+\infty }\left[ |u(x)|^{2}+|v(x)|^{2}\right] dx,
\label{P}
\end{equation}%
and the Hamiltonian (in which $\Omega =1$ is set),%
\begin{gather}
H=\int_{-\infty }^{+\infty }\left[ \frac{1}{2}\left( \left\vert \frac{%
\partial ^{2}u}{\partial x^{2}}\right\vert ^{2}+\left\vert \frac{\partial
^{2}v}{\partial x^{2}}\right\vert ^{2}\right) -\frac{\sigma }{2}\left(
|u|^{4}+|v|^{4}\right) \right.   \notag \\
\left. -\sigma g|u|^{2}|v|^{2}+\frac{1}{2}x^{2}\left( |u|^{2}+|v|^{2}\right)
-\lambda \left( uv^{\ast }+u^{\ast }v\right) \right] dx,  \label{H}
\end{gather}%
where $\ast $ stands for the complex conjugate.{\LARGE \ }The remaining
scaling invariance of Eq.~(\ref{uv}) makes it possible to either set $%
|\sigma |=1$, or keep nonlinearity coefficient $\sigma $ as a free
parameter, but fix $P\equiv 1$. All simulations performed in this work
comply with the conservation of $P$ and $H$, up to the accuracy of the
numerical codes.

It is relevant to mention that the present two-component system resembles
nonlinear models with a double-well potential, in the case when the wave
functions in two wells are linearly coupled by tunneling across the
potential barrier, see, e.g., Refs.~\cite{Oberthaler,double-well,Konotop}.
Nevertheless, the exact form of the system and its solutions are different.

In the limit of a small amplitude $A_{0}$ of the input, linearized equations
(\ref{uv}) with $\Omega =1$ admit exact solutions for inter-core JO of the
ground and dipole states (exact solutions for higher-order states can be
readily found too):%
\begin{eqnarray}
u_{\mathrm{JO}}^{\mathrm{(GS)}}\left( x,t\right) &=&A_{0}\exp \left( -\frac{1%
}{2}x^{2}-\frac{1}{2}iz\right) \cos \left( \lambda z\right) ,  \notag \\
&&  \label{exact0} \\
v_{\mathrm{JO}}^{\mathrm{(GS)}}\left( x,t\right) &=&iA_{0}\exp \left( -\frac{%
1}{2}x^{2}-\frac{1}{2}iz\right) \sin \left( \lambda z\right) ,  \notag
\end{eqnarray}%
\begin{eqnarray}
u_{\mathrm{JO}}^{\mathrm{(DM)}}\left( x,t\right) &=&A_{0}x\exp \left( -\frac{%
1}{2}x^{2}-\frac{3}{2}iz\right) \cos \left( \lambda z\right) ,  \notag \\
&&  \label{exact} \\
v_{\mathrm{JO}}^{\mathrm{(DM)}}\left( x,t\right) &=&iA_{0}x\exp \left( -%
\frac{1}{2}x^{2}-\frac{3}{2}iz\right) \sin \left( \lambda z\right) .  \notag
\end{eqnarray}%
Expressions given by Eqs. (\ref{exact0}) and (\ref{exact}) at $z=0$ are used
below as inputs in simulations of the full nonlinear equations (\ref{uv}).
The simulations presented in this section were performed in domain $|x|<10$.
This size tantamount to $20$ HO lengths is sufficient to display all details
of the solutions. Standard numerical methods were used, \textit{viz}., the
split-step fast-Fourier-transform scheme for simulations of the evolution
governed by Eqs. (\ref{uv}) and (\ref{asymm}), and the relaxation algorithm
for finding solutions of stationary equations, such as Eqs. (\ref{U}) and (%
\ref{V}), see below.

\subsection{The transition from regular to chaotic dynamics}

\begin{figure}[tbp]
\includegraphics[width=16cm]{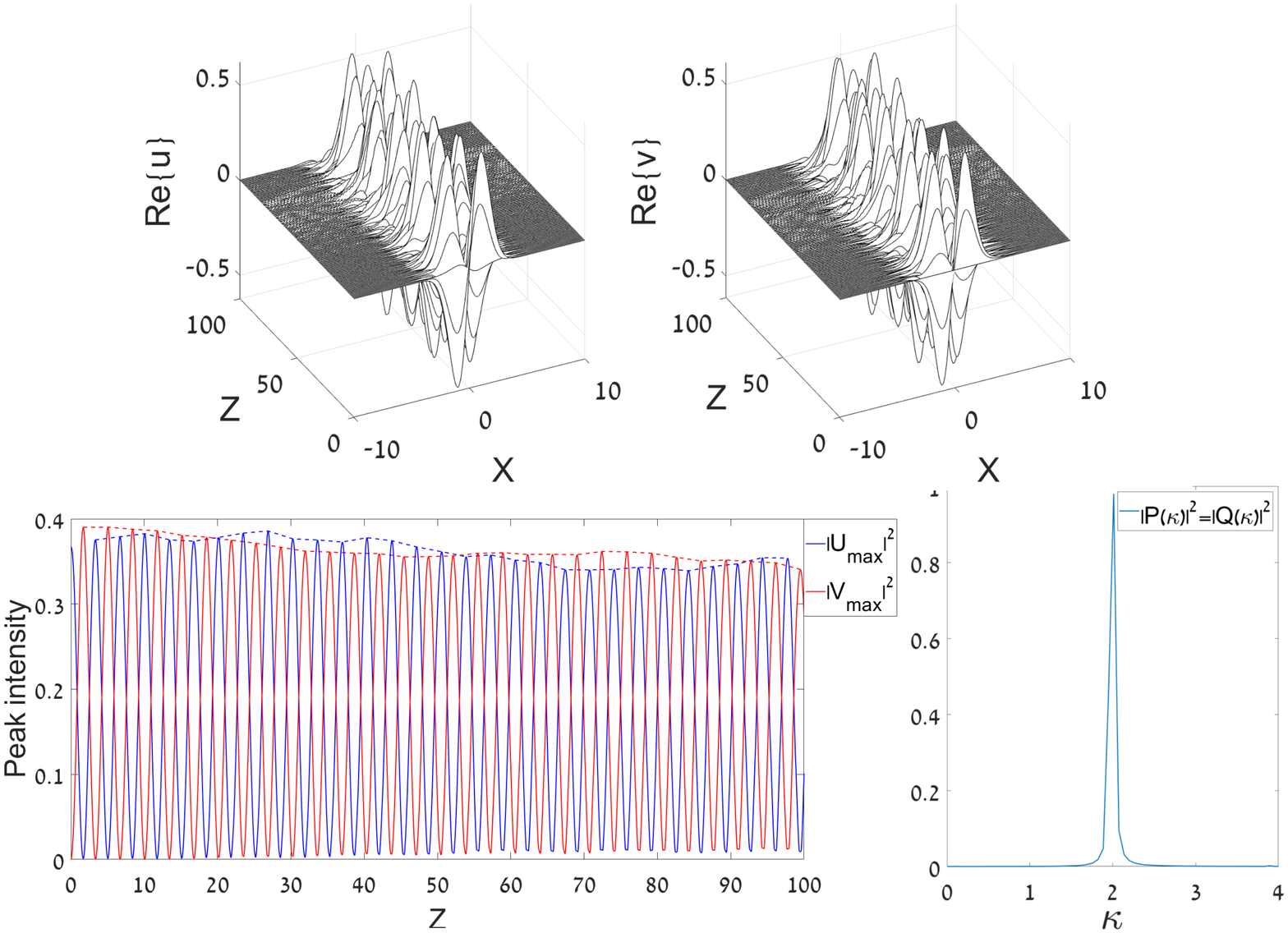}
\caption{A typical example of a regular Josephson dynamical regime,
initiated by the DM (dipole-mode) input launched in the $u$ component (as
given by Eq.~(\protect\ref{exact}) with $z=0$ and $A_{0}=1$). The solution
is produced by simulations of Eq.~(\protect\ref{uv}) with $\protect\lambda =%
\protect\sigma =\Omega =1$, $g=0$. Plots in the top row display the
evolution of components $u\left( x,z\right) $ and $v\left( x,z\right) $.
Left bottom: The evolution of the peak intensities of both components, $%
U_{\max }^{2}(z)\equiv \protect\underset{x}{\max }\left\{ \left\vert
u(x,z)\right\vert ^{2}\right\} $ and $V_{\max }^{2}(z)\equiv
\protect\underset{x}{\max }\left\{ \left[ \left\vert v(x,z)\right\vert ^{2}%
\right] \right\} $. Right bottom: The power spectrum of oscillations of the
two components, defined as per Eq.~(\protect\ref{Fourier}). The spectra are
virtually identical for both components.}
\label{fig1}
\end{figure}

Increase of amplitude $A_{0}$ of the input leads to nonlinear deformation of
the oscillations, and eventually to the onset of dynamical chaos. A typical
example of an essentially nonlinear but still regular JO dynamical regime,
produced by numerical simulations of Eq.~(\ref{uv}) with $g=0$ (no XPM\
interaction), in interval $0<z<Z$, with the DM input, taken as per Eq.~(\ref%
{exact}) at $z=0$, is presented in Fig. \ref{fig1}. In particular, the left
bottom panel of the figure displays oscillations of peak intensities of the
fields,
\begin{equation}
\left\{ U_{\max }^{2}(z),V_{\max }^{2}(z)\right\} \equiv \underset{x}{\max }%
\left\{ \left\vert u(x,z)\right\vert ^{2},\left\vert v(x,z)\right\vert
^{2}\right\} ,  \label{peak}
\end{equation}%
and, as a characteristic of the dynamics, in the right bottom panel we plot
power spectra, $\left\vert P(\kappa )\right\vert ^{2},|Q(\kappa )|^{2}$,
produced by the Fourier transform of the peak intensities:%
\begin{equation}
\left\{ P(\kappa ),Q(\kappa )\right\} =\int_{0}^{Z}\left\{ U_{\max
}^{2}(z),V_{\max }^{2}(z)\right\} \exp \left( -i\kappa z\right) dz,
\label{Fourier}
\end{equation}%
where $\kappa $ is a real propagation constant. Very slow decay of the peak
intensities, observed in the former panel, is a manifestation of the
system's nonintegrability (in this connection, we again stress that the
total norm is conserved in the course of the simulations).

The regularity of the dynamical regime displayed in Fig. \ref{fig1} is
clearly demonstrated by its spectral structure, which exhibits a single
narrow peak at $\kappa _{\mathrm{peak}}\approx 2$. The peak's width, $\Delta
\kappa \simeq 0.1$, which corresponds to the relative width, $\Delta \kappa
/\kappa _{\mathrm{peak}}\approx 0.05$, is comparable to the spread of the
Fourier transform, corresponding to $Z=100$ in Fig. \ref{fig1}. It can be
estimated as $\delta \kappa =2\pi /Z\approx 0.06$. Note the overall symmetry
between the two components in Fig. \ref{fig1} (in particular, their spectra
are identical in the bottom panel).

\begin{figure}[tbp]
\includegraphics[width=16cm]{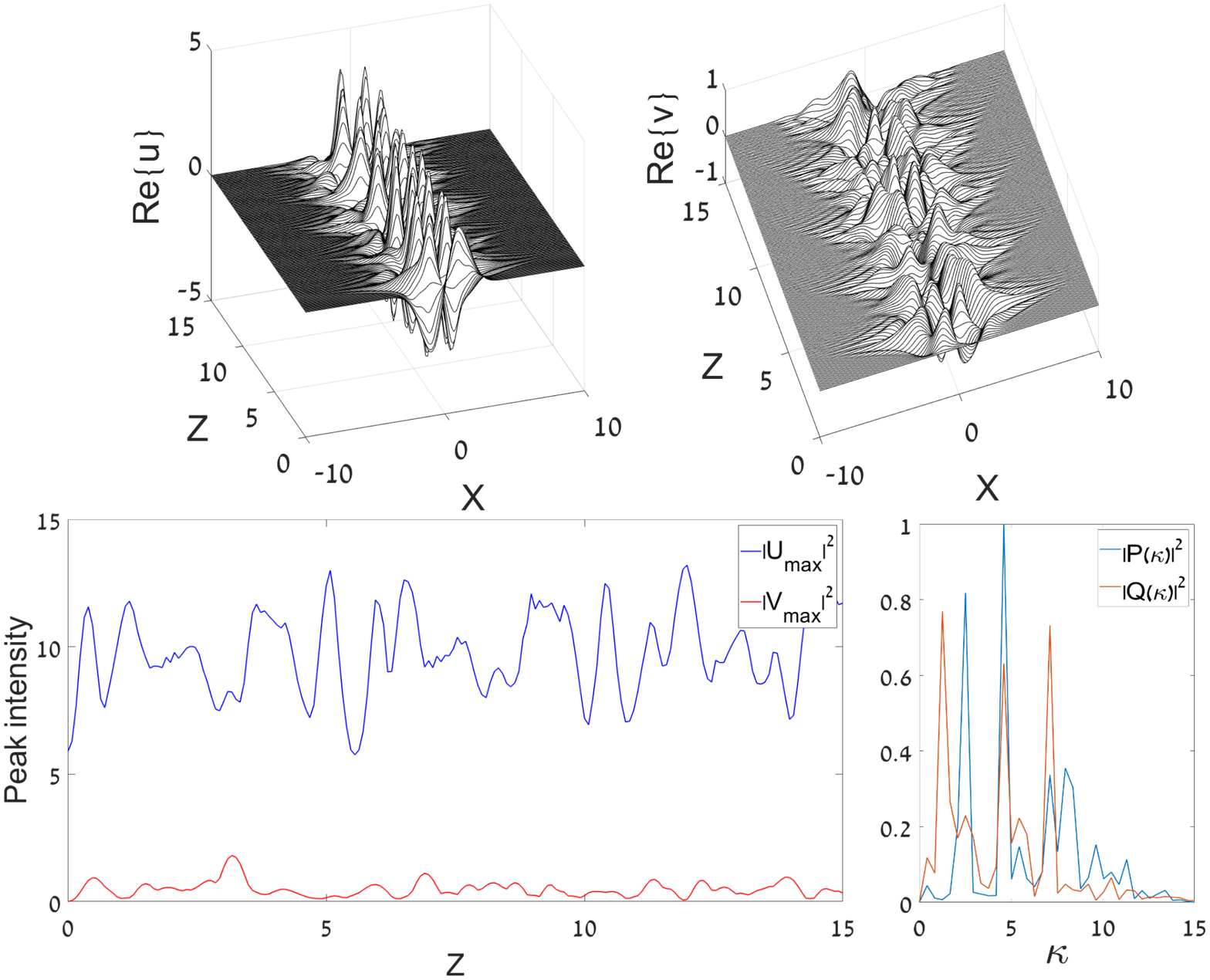}
\caption{The same as in Fig. \protect\ref{fig1}, but for a typical example
of a chaotic dynamical regime, initiated by the DM input (\protect\ref{exact}%
) with a larger amplitude, $A_{0}=4$. Note different scales on vertical axes
in two plots in the left panel. }
\label{fig2}
\end{figure}

The simulations with the same input, but larger values of $A_{0}$, give rise
to chaotic (``turbulent") dynamical states with a broad dynamical spectrum,
see a typical example in Fig. \ref{fig2}. Note that both the regular and
chaotic dynamical pictures displayed in Figs. \ref{fig1} and \ref{fig2}
extend over the distance estimated to be $\symbol{126}10$ Rayleigh
(diffraction) lengths corresponding to the width of the DM input. This
estimate is sufficient to make conclusions about the character of the
dynamics.

Systematic simulations with the GS input, provided by Eq.~(\ref{exact0}) at $%
z=0$, produce similar results (not shown here in detail). In particular, as
well as in the case of the DM input, amplitudes $A_{0}=1$ and $4$ initiate,
severally, quasi-regular and chaotic evolution.

\begin{figure}[tbp]
\includegraphics[width=16cm]{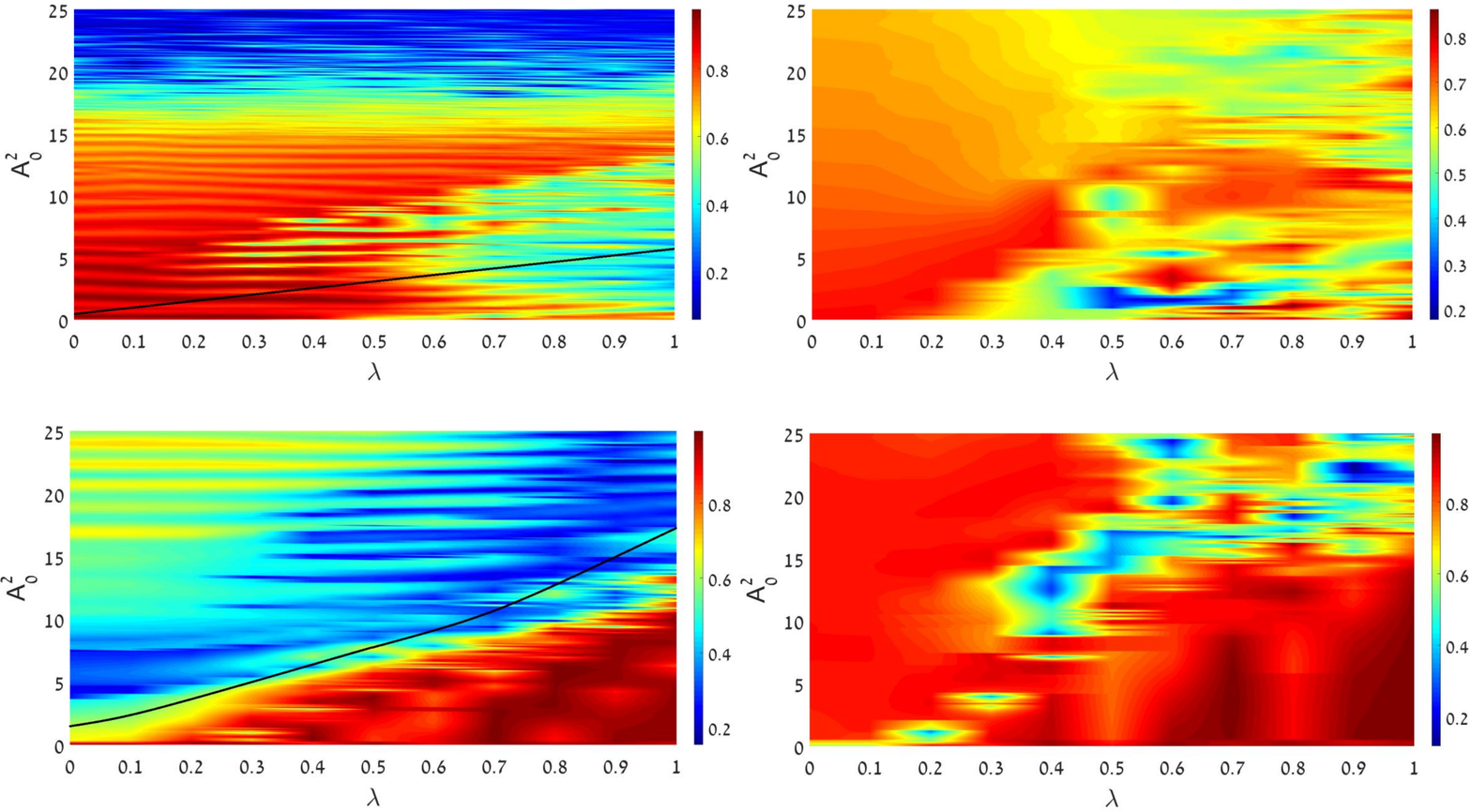}
\caption{Heatmaps of values of sharpness (\protect\ref{Sh}) of the central
spectral peak quantifying proximity of the system's dynamics to the regular
regime. The maps are plotted in the plane of the linear-coupling strength, $%
\protect\lambda $, and intensity of the input, $A_{0}^{2}$, which is
launched in one component. Top left: The GS input, given by Eq.~(\protect\ref%
{exact0}) at $z=0$, in the case of the self-attraction ($\protect\sigma =+1$%
). Top right: The same, but in the case of self-repulsion ($\protect\sigma %
=-1$). Bottom left: The same as in the the top left panel, but produced by
the DM input, given by Eq.~(\protect\ref{exact}) at $z=0$. Bottom right: The
same as in the bottom left panel, but in the case of self-repulsion ($%
\protect\sigma =-1$). In all cases, $g=0$ is set in Eq.~(\protect\ref{uv})
(no XPM interaction between the components). Black curves cutting the left
panels in their lower areas designate the onset of SSB (spontaneous symmetry
breaking), signalized by appearance of $\protect\theta \neq 0$, see Eq.~(%
\protect\ref{theta}).}
\label{fig3}
\end{figure}

The results for the transition from regular JO to chaotic dynamics,
initiated by the GS and DM inputs, taken as per Eqs. (\ref{exact0}) and (\ref%
{exact}) at $z=0$, are summarized in charts plotted in the plane of $\left(
\lambda ,A_{0}^{2}\right) $ in Fig. \ref{fig3}. They display heatmaps of
values of the parameter which quantifies the sharpness of the central peak
in the spectrum of the dynamical state:%
\begin{equation}
\mathrm{Sharpness}\equiv \frac{\int_{\mathrm{FWHM}}\left\vert P(\kappa
)\right\vert ^{2}d\kappa }{\int_{0}^{\infty }\left\vert P(\kappa
)\right\vert ^{2}d\kappa },  \label{Sh}
\end{equation}%
where the integration in the numerator is performed over the section of the
central spectral peak selected according to the standard definition of the
full width at half-maximum: $\left\vert P(\kappa _{\mathrm{FWHM}%
})\right\vert ^{2}=(1/2)\left( P(\kappa )\right) _{\max }$. Values of $%
\mathrm{Sharpness}$ close to $1$ imply the domination of a single sharp
peak, such as one in Fig. \ref{fig1}, which corresponds to a regular
dynamical regime, while decrease of this parameter indicates a transition to
a broad spectrum, which is a telltale of the onset of chaotic dynamics --
see, e.g., Fig. \ref{fig2}.

Figure \ref{fig3} clearly demonstrates decay of the central-peak's
sharpness, i.e., transition to dynamical chaos, with the increase of the
input's intensity, $A_{0}^{2}$, in the case of the attractive SPM, $\sigma
=+1$. Such a trend is not straightforward in the opposite case of the
self-repulsion, $\sigma =-1$. In particular, the chaotization is not
observed at all in the latter case at small values of $\lambda $. This
conclusion agrees with findings reported in work \cite{Newcastle} for the
single-component NLSE (which corresponds to $\lambda =0$), with the HO
potential and $\sigma =-1$. Computations of the spectrum, reported in that
work, demonstrate that no transition to dynamical chaos takes place at all
values of parameters. The fact that an area of weak chaotization is,
nevertheless, observed in the right panels of Fig. \ref{fig3} is explained
by the above-mentioned circumstance, that the addition of the linear
coupling to a pair of NLSEs destroys their integrability (in free space). On
the other hand, the increase of the sharpness with the increase of $\lambda $%
, observed at relatively small values of $A_{0}^{2}$ (which is observed in
an especially salient form in the left bottom panel of Fig. \ref{fig3}) also
has a simple explanation: the increase of $\lambda $ makes the linear terms
in the system dominating over nonlinear ones, thus tending to maintain a
quasi-linear behavior.

Lastly, the comparison of the left and right panels in Fig. \ref{fig3}
suggests that the chaotization sets in faster in dynamical regimes initiated
by the DM input, in comparison to their counterparts originating from the
GS, for the same values of the input's intensity, $A_{0}^{2}$. The
difference between the GS and DM dynamical regimes is salient for relatively
small values of $\lambda $. It may be explained by the fact that attractive
SPM naturally tends to form a stable bright soliton from the GS input, which
then maintains regular motion in the HO potential \cite{RMP}. On the other
hand, spatially odd bright solitons do not exist in free space, which
impedes transformation of the DM input into a regular dynamical state.

As said above, the heatmaps are displayed in Fig. \ref{fig3} for $g=0$,
i.e., in the absence of the XPM coupling between the components, which is
the case for dual-core couplers. On the other hand, in the case of the
Manakov's nonlinearity, i.e., $g=1$, the above-mentioned integrability of
such a system of NLSEs with the linear coupling \cite{ST} (but without the
trapping potential) suggests that the full system will be closer to
integrability and farther from the onset of chaos. Indeed, numerical results
collected from simulations of Eq.~(\ref{uv}) with $\sigma =g=+1$ (not shown
here in detail) demonstrate a much smaller chaotic area in the $\left(
\lambda ,A_{0}^{2}\right) $ plane. In particular, the GS input generates
``turbulent" behavior only at $A_{0}^{2}\gtrsim 200$, being limited to $%
\lambda \lesssim 0.15$, cf. Fig. \ref{fig3}(a).

\subsection{Spontaneous symmetry breaking (SSB) between the coupled
components}

A noteworthy feature of the dynamical state presented in Fig. \ref{fig2} is
breaking of the symmetry between fields $u$ and $v$ (while the patterns
initiated by the GS and DM inputs keep their parities, i.e., spatial
symmetry (evenness) and antisymmetry (oddness), respectively). This is a
manifestation of the general effect which is well known, in diverse forms,
in linearly coupled dual-core systems with intrinsic attractive SPM \cite%
{book}. In particular, SSB of stationary states in systems with the HO
trapping potential acting in both cores was addressed in Refs.~\cite{Viskol}
and \cite{HS2}, while Fig. \ref{fig2} demonstrates the symmetry breaking in
the \emph{dynamical} JO state.

The SSB effect may be quantified, as usual \cite{OptJO6}, by asymmetry of
the dynamical states, which we define as%
\begin{equation}
\Theta \equiv \frac{\int_{0}^{\infty }\left\vert P(\kappa )\right\vert
^{2}d\kappa -\int_{0}^{\infty }\left\vert Q(\kappa )\right\vert ^{2}d\kappa
}{\int_{0}^{\infty }\left\vert P(\kappa )\right\vert ^{2}d\kappa
+\int_{0}^{\infty }\left\vert Q(\kappa )\right\vert ^{2}d\kappa }.
\label{theta}
\end{equation}%
The SSB occurs as a transition from $\Theta =0$ to $\Theta \neq 0$ with the
increase of $A_{0}^{2}$ at some critical point, which is a generic property
of stationary states in dual-core systems with intrinsic attractive
nonlinearity \cite{book,OptJO6}, while here we consider it in the dynamical
setting. On the other hand, in the case of the repulsive nonlinearity, $%
\sigma =-1$ in Eq.~(\ref{uv}), SSB of stationary states takes place in this
system only at $g>1$ \cite{HS2}, while we here focus on the most relevant
case of $g=0$. Accordingly, the present system with $\sigma =-1$ does not
feature SSB.

\begin{figure}[tbp]
\includegraphics[width=16cm]{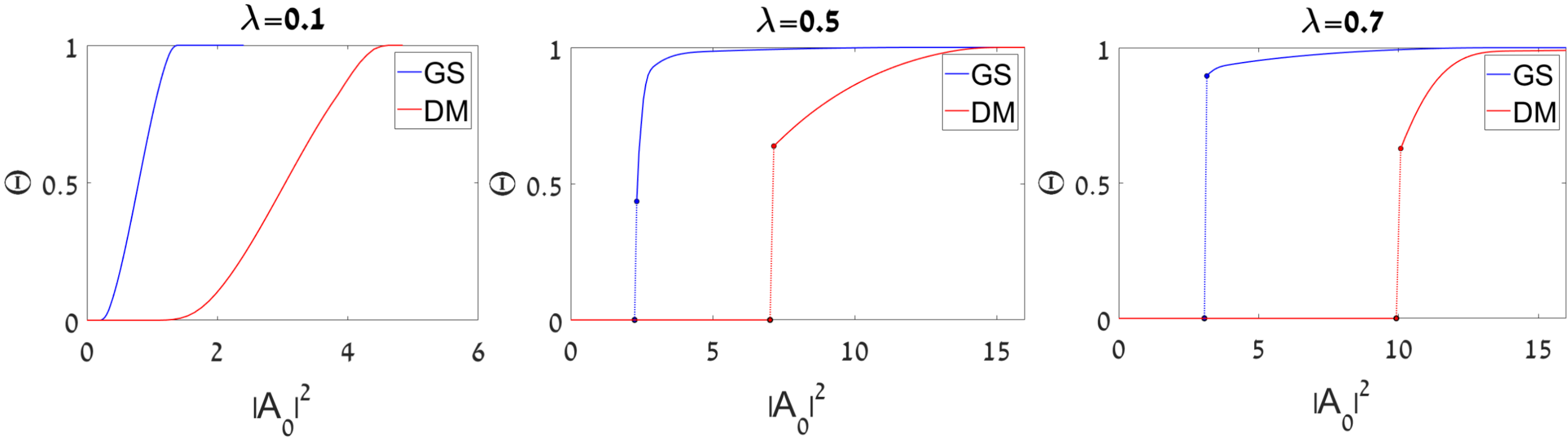}
\caption{{}Plots of the SSB (spontaneous symmetry breaking) in the dynamical
states initiated by the GS (ground-state) and DM (dipole-mode) inputs: the
asymmetry parameter, defined as per Eq.~(\protect\ref{theta}), is plotted
versus the intensity of the input, $A_{0}^{2}$, for three different fixed
values of the linear-coupling constant, $\protect\lambda $, as indicated in
the figure.}
\label{fig4}
\end{figure}

For $\sigma =+1$, the SSB boundaries in the parameter planes of the GS and
DM solutions are shown by bold black lines in the top and bottom left panels
of Figs. \ref{fig3}, respectively. The SSB bifurcation of the dynamical
states under the consideration is of the \textit{supercritical}, alias
\textit{forward}, type \cite{Iooss}, in terms of dependence $\Theta \left(
A_{0}^{2}\right) ,$ as shown in Fig. \ref{fig4} for the dynamical states
initiated by the GS and DM inputs. It is observed that, naturally, the
critical value of $A_{0}^{2}$ increases with $\lambda $, as the symmetry is
maintained by the linear coupling, hence stronger coupling needs stronger
nonlinearity to break the symmetry. Note also that the transition from $%
\Theta =0$ to $\Theta \approx 1$ (a strongly asymmetric state) is steeper at
larger $\lambda $.

It is worthy to note that, as clearly shown by the SSB boundary (the black
line) in the top left panel of Fig. \ref{fig3}, the SSB of the GS mode
happens prior to the transition to the dynamical chaos. This conclusion
agrees with known results showing that SSB of stationary modes does not,
normally, lead to chaotization of the system's dynamics \cite{book,OptJO6}.
On the other hand, the bottom left panel in Fig. \ref{fig3} demonstrates a
different situation for the DM states, which exhibit the chaotization prior
to the SSB, although the separation between these transitions is small. This
conclusion agrees with the fact that, as clearly seen in Fig. \ref{fig4},
the SSB in DM\ states occurs at values of the input's amplitude essentially
higher than those which determine the SSB threshold of the GS solutions.

\section{The half-trapped system}

The asymmetric system of linearly-coupled NLSEs, with the HO potential
included in one equation only, is written as%
\begin{eqnarray}
iu_{z}+\frac{1}{2}u_{xx}+\lambda v-\frac{1}{2}x^{2}u+\sigma \left(
|u|^{2}+g|v|^{2}\right) u &=&-\omega u,  \notag \\
&&  \label{asymm} \\
iv_{z}+\frac{1}{2}v_{xx}+\lambda u+\sigma \left( |v|^{2}+g|u|^{2}\right) v
&=&0,  \notag
\end{eqnarray}%
cf. Eq.~(\ref{uv}). Here, as said above, $\Omega =1$ is set in the first
equation, and the propagation-constant mismatch, $\omega $ (in terms of BEC,
it represents a difference in the chemical potentials between the two wave
functions) is a common feature of asymmetric systems. Stationary solutions
to Eq.~(\ref{asymm}) are looked for as%
\begin{equation}
\left\{ u,v\right\} =\left\{ U(x),V(x)\right\} \exp \left( -i\mu z\right) ,
\label{uvUV}
\end{equation}%
with real propagation constant $-\mu $ (in BEC, with $z$ replaced by $t$, $%
\mu $ is the chemical potential), and real functions $U(x)$ and $V(x)$
satisfying equations%
\begin{gather}
\left( \mu +\omega \right) U+\frac{1}{2}\frac{d^{2}U}{dx^{2}}+\lambda V-%
\frac{1}{2}x^{2}U+\sigma \left( U^{2}+gV^{2}\right) U=0,  \label{U} \\
\mu V+\frac{1}{2}\frac{d^{2}V}{dx^{2}}+\lambda U+\sigma \left(
V^{2}+gU^{2}\right) V=0.  \label{V}
\end{gather}%
Most results are produced below disregarding the XPM coupling between the
components ($g=0$). Nevertheless, the XPM terms are included when collecting
numerical results for the threshold of the existence of bound states,
displayed in Fig. \ref{fig14}.

\subsection{The linearized system: analytical and numerical results}

\subsubsection{Emission of radiation in the untrapped component}

In the linear limit, $\sigma =0$, two decoupled equations in system (\ref%
{asymm}) with $\lambda =0$ produce completely different results: all
excitations of component $u$ stay confined in the HO trap, while the $v$
component with any $\mu >-\omega $ freely expands. In particular, in the
limit of $\lambda =0$, obvious bound-state GS and DM\ solutions of the
linearized version of Eqs. (\ref{U}) and (\ref{V}), with zero $v$ component,
are \ \
\begin{eqnarray}
U_{\mathrm{GS}}^{(0)}(x) &=&\frac{1}{\pi ^{1/4}}\exp \left( -\frac{x^{2}}{2}%
\right) ,V_{\mathrm{GS}}^{(0)}=0,  \label{00} \\
U_{\mathrm{DM}}^{(0)}(x) &=&\frac{\sqrt{2}}{\pi ^{1/4}}x\exp \left( -\frac{%
x^{2}}{2}\right) ,V_{\mathrm{DM}}^{(0)}=0,  \label{10}
\end{eqnarray}%
where the pre-exponential constants are determined by the normalization
condition,
\begin{equation}
\int_{-\infty }^{+\infty }\left[ U^{2}(x)+V^{2}(x)\right] dx=1,  \label{P=1}
\end{equation}%
which we adopt in this section. The eigenvalues corresponding to eigenmodes (%
\ref{00}) and (\ref{10}) are%
\begin{equation}
\mu _{\mathrm{GS}}^{(0)}=1/2-\omega ;~\mu _{\mathrm{DM}}^{(0)}=3/2-\omega .
\label{mu0}
\end{equation}

Proceeding to dynamical states, in the lowest approximation with respect to
small $\lambda $ the evolution of the $v$ field is driven by the respective
linearized equation in system (\ref{asymm}),%
\begin{equation}
iv_{z}+\frac{1}{2}v_{xx}=-\lambda U_{\mathrm{GS,DM}}^{(0)}(x)\exp \left(
-i\mu _{\mathrm{GS,DM}}^{(0)}z\right) .  \label{vU}
\end{equation}%
Obviously, Eq.~(\ref{vU}) gives rise to emission of propagating waves
(``radiation"), in the form of $v\sim \lambda \exp \left( ikx-i\mu _{\mathrm{%
GS,DM}}^{(0)}z\right) $, at resonant wavenumbers $k=\pm \sqrt{2\mu _{\mathrm{%
GS,DM}}^{(0)}}$, provided that $\mu _{\mathrm{GS,DM}}^{(0)}$ is positive,
i.e.,%
\begin{equation}
v_{\mathrm{rad}}\sim \lambda \exp \left( \pm i\sqrt{2\mu _{\mathrm{GS,DM}%
}^{(0)}}\left( x-V_{\mathrm{ph}}z\right) \right) ,  \label{emission}
\end{equation}%
where the phase velocity is%
\begin{equation}
V_{\mathrm{ph}}=\frac{k}{2}\equiv \pm \frac{1}{2}\sqrt{2\mu _{\mathrm{GS,DM}%
}^{(0)}}  \label{Vph}
\end{equation}%
(in terms of the spatial-domain propagation in the optical waveguide, it is
actually the beam's slope). The expansion of the area in the $\left(
x,z\right) $ plane occupied by the radiation field is bounded by the group
velocity, $\left\vert V_{\mathrm{gr}}\right\vert =|k|\equiv 2V_{\mathrm{ph}}$%
.

\begin{figure}[tbp]
\includegraphics[width=16cm]{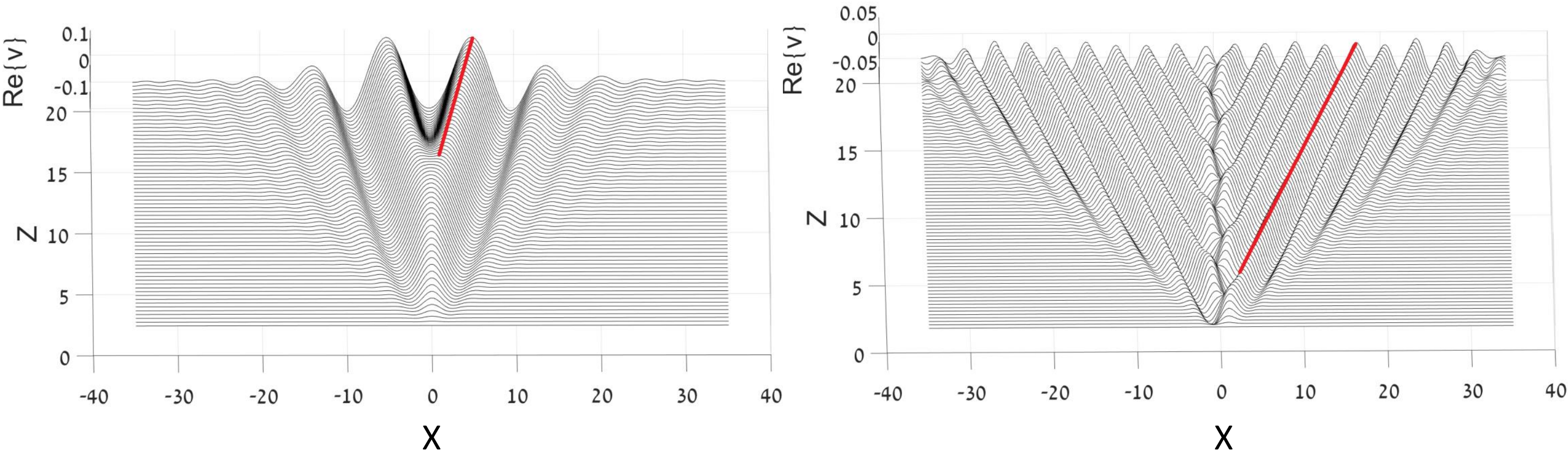}
\caption{{}Simulations of the evolution of the linearized half-trapped
system (\protect\ref{asymm}), displayed in the untrapped component by
plotting Re$\left( v\left( x,z\right) \right) $. Left: Emission of radiation
generated by the GS (ground-state) populating the trapped component, $%
u\left( x,z\right) $ (see Eq.~(\protect\ref{00})), with $\protect\omega %
=0.25 $ in Eq.~(\protect\ref{asymm}). Right: The same, but for the radiation
generated by the DM (dipole mode) in the trapped component (see Eq.~(\protect
\ref{10})), with $\protect\omega =0$. In both cases, the linear-coupling
constant is $\protect\lambda =0.05$.}
\label{fig5}
\end{figure}

An illustration of this dynamics is presented in Fig. \ref{fig5}. Straight
red lines designate the wave-propagation directions, which exactly agree
with the phase velocity predicted by Eq.~(\ref{Vph}), and the expansion of
the area occupied by the radiation complies with the prediction based on the
group velocity.

\begin{figure}[tbp]
\includegraphics[width=16cm]{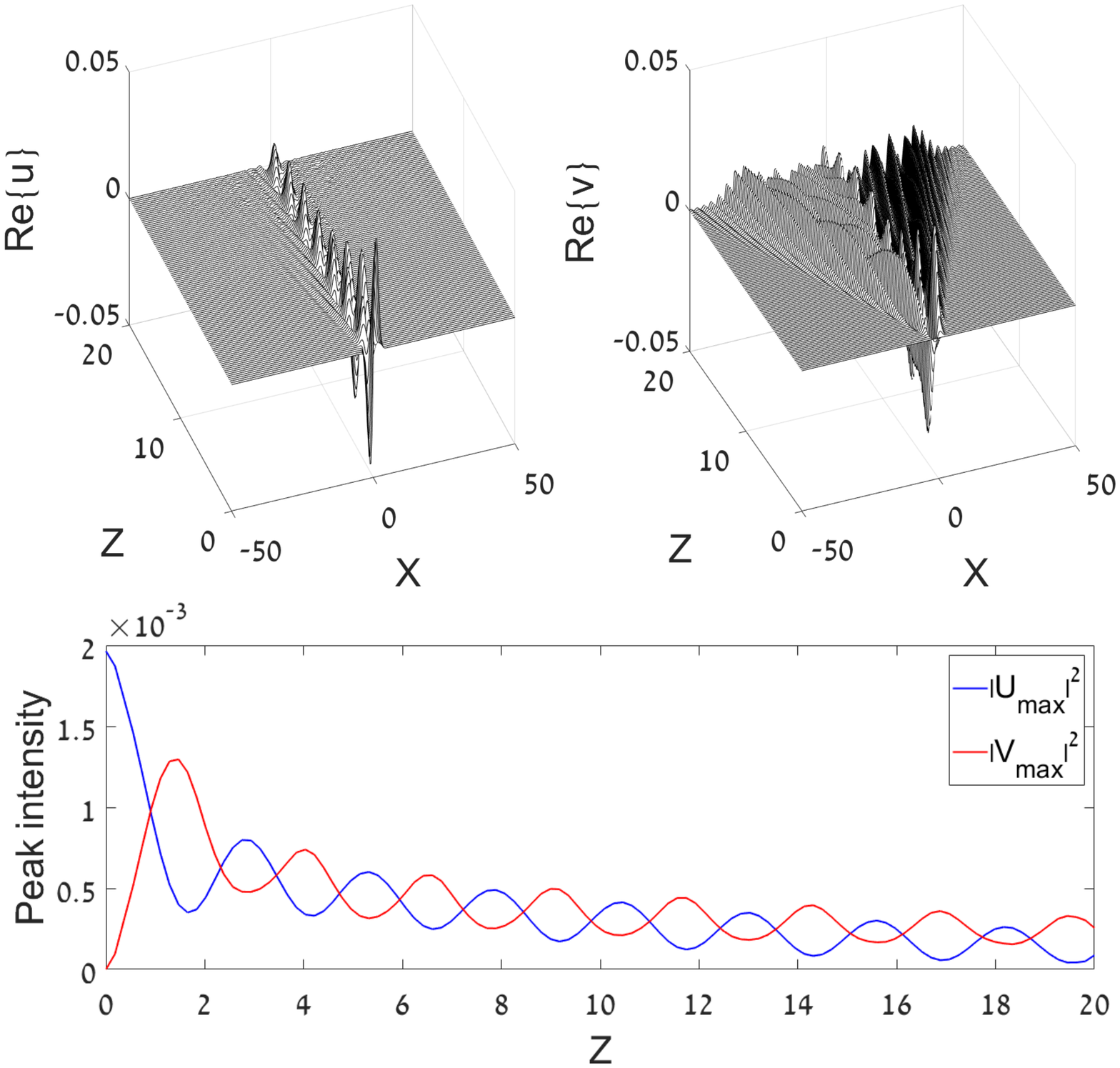}
\caption{Left: The evolution of fields $u$ and $v$ in the half-trapped
system, as produced by simulations of Eq.~(\protect\ref{asymm}) with $%
\protect\omega =0$ and coupling constant $\protect\lambda =1$, initiated by
the DM\ input in the $u$ core, taken as per Eq.~(\protect\ref{00}) with $%
A_{0}=0.1$. Here and in Fig. \protect\ref{fig7}, spurious left-right
asymmetry of the radiation field in the $v$ component is an illusion
produced by plotting. Right: The evolution of the peak intensities of both
components, $\protect\underset{x}{\max }\left\{ \left\vert u(x,t)\right\vert
^{2}\right\} $ and $\protect\underset{x}{\max }\left\{ \left\vert v\left(
x,t\right) \right\vert ^{2}\right\} $.}
\label{fig6}
\end{figure}

The emission of radiation into the $v$ core gives rise to a gradual decay of
the amplitude in the $u$ core, due to the conservation of the total norm,
see Eq.~(\ref{P=1}). An example of the decay is displayed in Fig. \ref{fig6}%
, for a small initial amplitude of the GS input, $A_{0}=0.1$ (which
corresponds to the quasi-linear dynamical regime), and a relatively large
coupling constant, $\lambda =1$, which makes the transfer of the norm
(power) from $u$ to $v$ faster.
On the other hand, the same input with large $A_{0}$ makes the $u$-$v$
coupling a weak effect, in comparison with the dominant nonlinearity, hence
the input mode in the $u$ core seems quite stable, as shown in Fig. \ref%
{fig7}.
\begin{figure}[tbp]
\includegraphics[width=16cm]{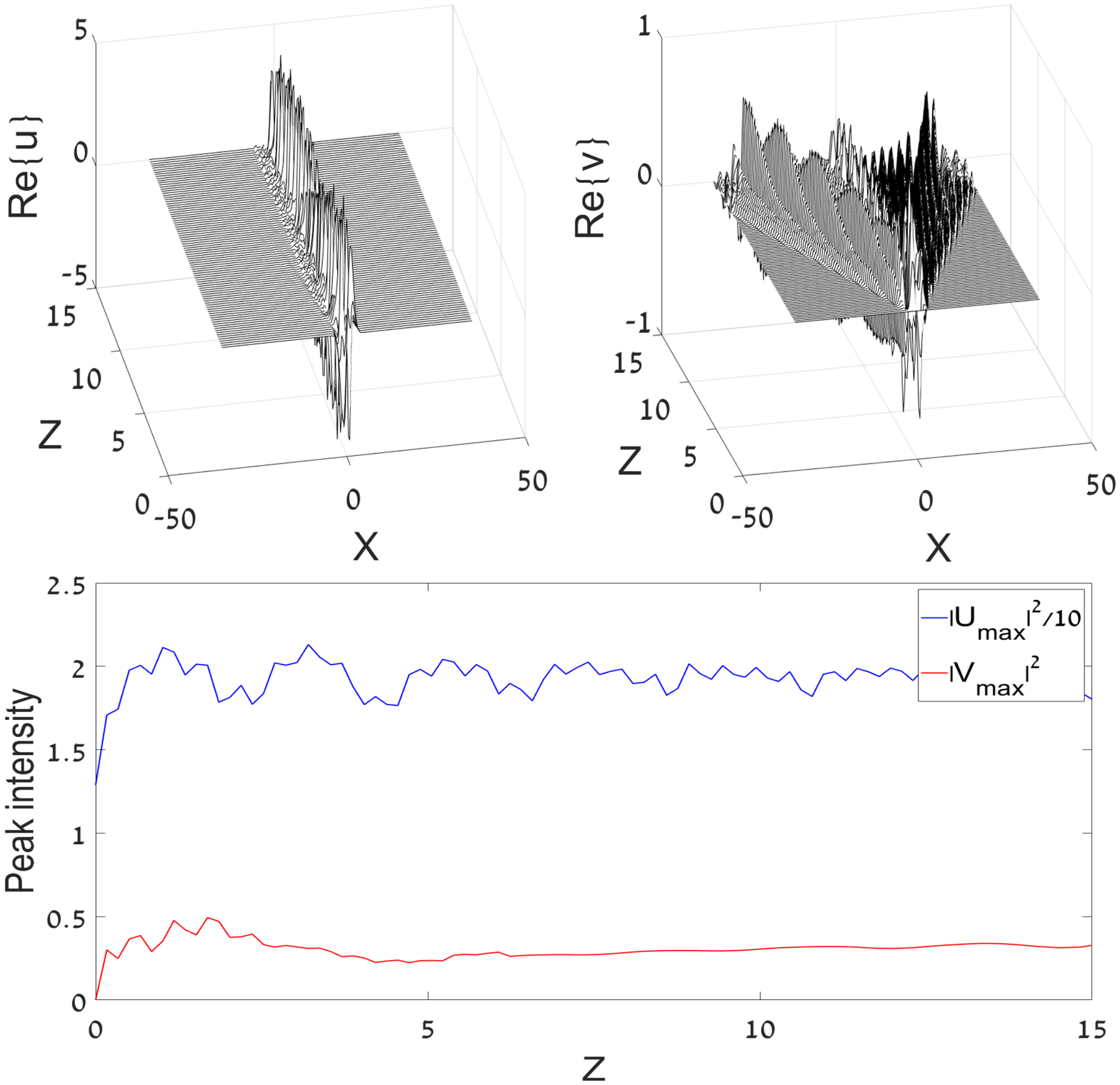}
\caption{{}The same as in Fig. \protect\ref{fig6}, but for a large amplitude
of the DM input in the $u$ component, $A_{0}=8$.}
\label{fig7}
\end{figure}

\subsubsection{The shift of the GS and DM existence thresholds at small
values of coupling constant $\protect\lambda $}

At $\mu _{\mathrm{GS,DM}}^{(0)}<0$ the radiation is not generated by Eq.~(%
\ref{emission}), hence two-component bound states may exist in this case.
Our next objective is to find, for the coupled system with $\lambda >0$,
threshold values $\left( \omega _{\mathrm{GS,DM}}\right) _{\mathrm{thr}}$ of
the mismatch parameter $\omega $ in Eqs. (\ref{U}) and (\ref{V}), such that
the bound states of the GS and DM types exist at
\begin{equation}
\omega >\left( \omega _{\mathrm{GS,DM}}\right) _{\mathrm{thr}},  \label{thr}
\end{equation}%
respectively. Obviously, $\left( \omega _{\mathrm{GS}}\right) _{\mathrm{thr}%
}=1/2$ and $\left( \omega _{\mathrm{DM}}\right) _{\mathrm{thr}}=3/2$ in the
limit of $\lambda =0$, see Eq.~(\ref{mu0}).

First, we aim to find lowest-order corrections to the GS and DM eigenvalues (%
\ref{mu0}) for small $\lambda $. Then, a shift of the respective thresholds
can be identified by setting $\mu _{\mathrm{GS,DM}}=0$. In the limit of $%
\lambda =0$, the GS and DM wave functions are taken as per Eqs. (\ref{00})
and (\ref{10}), respectively. With small $\lambda $, the first-order
solution for $V(x)$, \textit{viz}., $V(x)\equiv \lambda V_{\mathrm{GS,DM}%
}^{(1)}(x)$, has to be found from the inhomogeneous equation, that follows
from Eq.~(\ref{V}), in which $\mu =0$ is set:
\begin{equation}
\frac{d^{2}}{dx^{2}}V_{\mathrm{GS,DM}}^{(1)}=-2U_{\mathrm{GS,DM}}^{(0)}(x).
\label{V1}
\end{equation}%
Straightforward integration of Eq.~(\ref{V1}), with expressions (\ref{00})
and (\ref{10}) substituted on the right-hand side, yields%
\begin{equation}
V_{\mathrm{GS}}^{(1)}(x)=-\sqrt{2}\pi ^{1/4}\left[ \frac{x}{\sqrt{2}}\mathrm{%
erf}\left( \frac{x}{\sqrt{2}}\right) +\sqrt{\frac{2}{\pi }}\exp \left( -%
\frac{x^{2}}{2}\right) \right] ,  \label{V01}
\end{equation}%
\begin{equation}
V_{\mathrm{DM}}^{(1)}(x)=2\pi ^{1/4}\mathrm{erf}\left( \frac{x}{\sqrt{2}}%
\right) ,  \label{V11}
\end{equation}%
where $\mathrm{erf}(x)$ is the standard error function, which is an odd
function of $x$.

Next, the small perturbation in Eq.~(\ref{U}), represented by term $\lambda
V $, produces a small shift $\delta \mu $ of the eigenvalue, as a feedback
from component $V$. According to the standard rule of quantum mechanics,
which deals with the linear Schr\"{o}dinger equation \cite{LL}, in the first
approximation of the perturbation theory, when $V(x)$ is replaced by
expression (\ref{V01}) or (\ref{V11}), the result is%
\begin{equation}
\delta \mu _{\mathrm{GS,DM}}=-\lambda ^{2}\int_{-\infty }^{+\infty }V_{%
\mathrm{GS,DM}}^{(1)}(x)U_{\mathrm{GS,DM}}^{(0)}(x)dx\equiv -I_{\mathrm{GS,DM%
}}\lambda ^{2},  \label{delta}
\end{equation}%
with coefficients
\begin{equation}
I_{\mathrm{GS}}=-3.414,I_{\mathrm{DM}}=4,~  \label{II}
\end{equation}%
where the former and latter ones are computed, respectively, in a numerical
form and analytically (note \emph{opposite signs} of these coefficients).
Then, the accordingly shifted threshold values sought for are%
\begin{equation}
\left( \omega _{\mathrm{GS}}\right) _{\mathrm{thr}}=1/2-I_{\mathrm{GS}%
}\lambda ^{2},~\left( \omega _{\mathrm{DM}}\right) _{\mathrm{thr}}=3/2-I_{%
\mathrm{DM}}\lambda ^{2}.  \label{smalllambda}
\end{equation}

\begin{figure}[tbp]
\includegraphics[width=16cm]{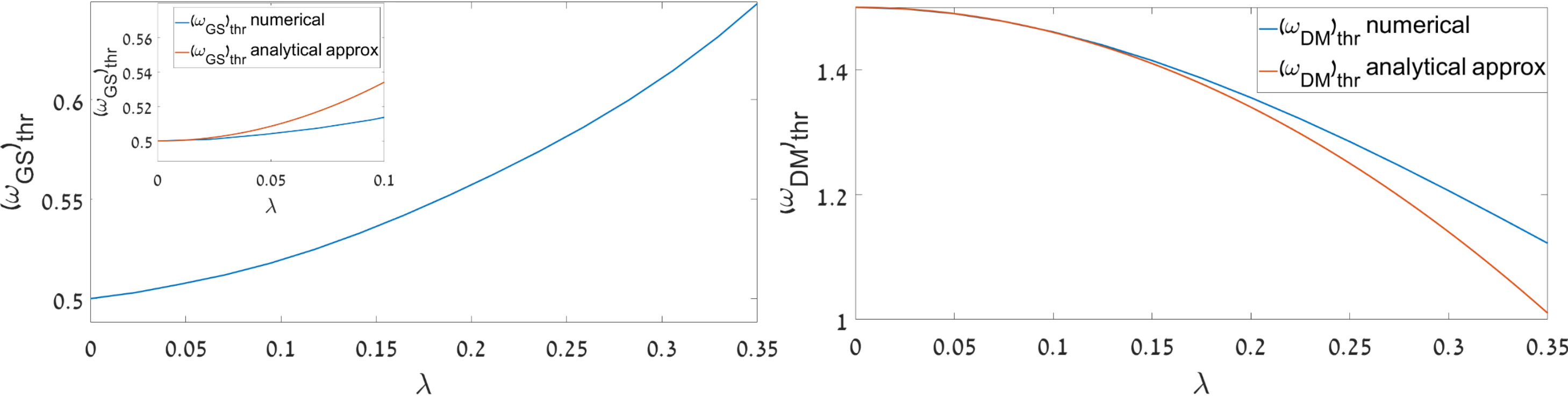}
\caption{The analytically predicted and numerically found threshold values
of the mismatch parameter, $\protect\omega $, above which the GS and DM
solutions (the left and right panels, respectively) are produced, for the
half-trapped system, by the linearized version of Eqs. (\protect\ref{U}) and
(\protect\ref{V}), vs. coupling constant $\protect\lambda $. The analytical
results are produced by Eq.~(\protect\ref{smalllambda}). For the GS, they
are shown (in the inset) only for relatively small values of $\protect%
\lambda $, as in this case the analytical approximation is inaccurate at
larger $\protect\lambda $.}
\label{fig8}
\end{figure}

\begin{figure}[tbp]
\includegraphics[width=12.0cm]{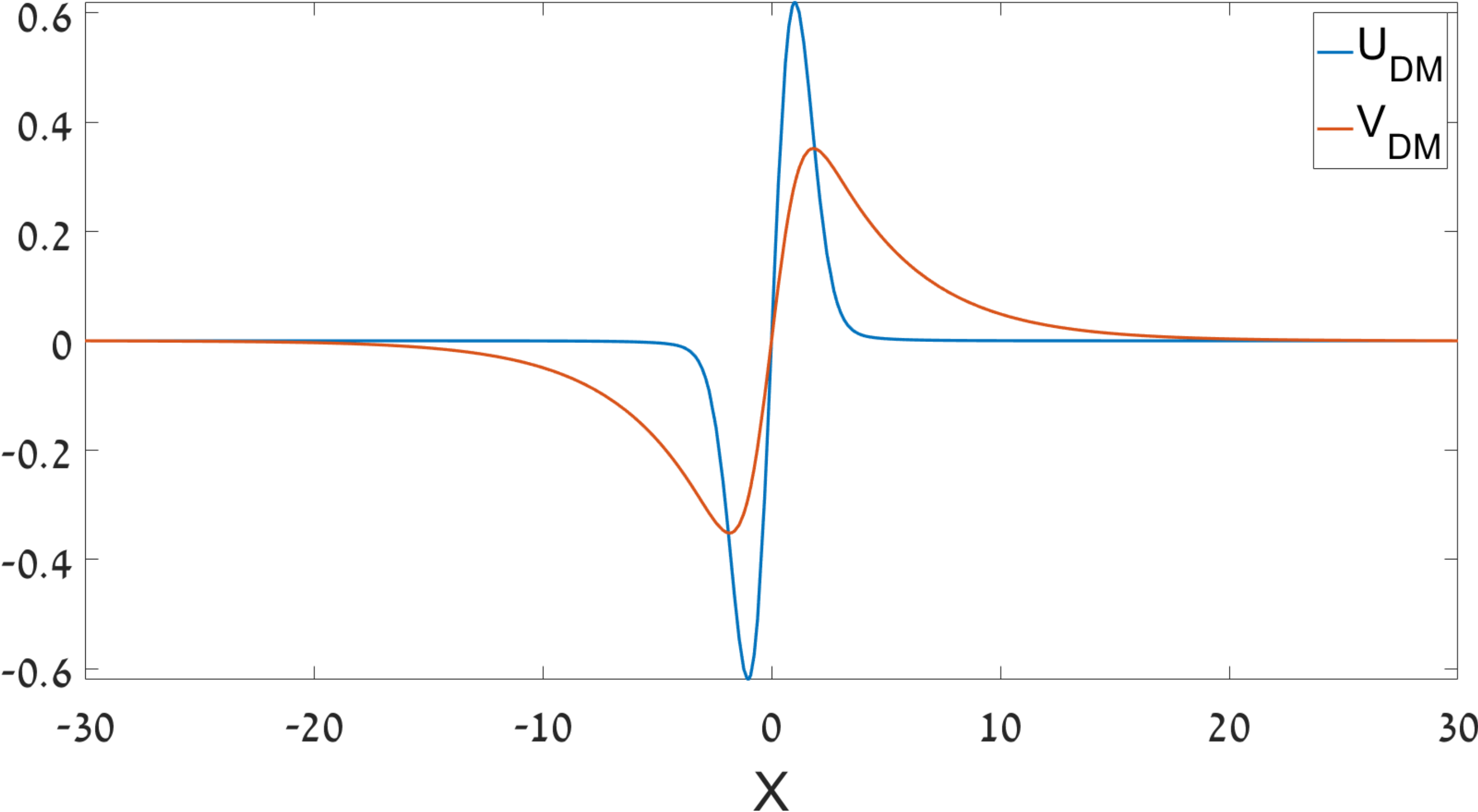}
\caption{A bound state of the DM (dipole-mode) type in the half-trapped
system, found as a numerical solution of Eqs. (\protect\ref{U}) and (\protect
\ref{V}) with $\protect\omega =1.4$, $\protect\lambda =0.225$, and $\protect%
\sigma =0$ (the linearized version). The eigenvalue corresponding to this
solution is $\protect\mu \approx -0.036$. }
\label{fig9}
\end{figure}

The analytical predictions are compared to numerical results, obtained from
a solution of the linearized variant of Eqs. (\ref{U}) and (\ref{V}), in
Fig. \ref{fig8}. Note that the linear $u$-$v$ coupling \emph{facilitates}
the formation of the DM bound state, by lowering $\left( \omega _{\mathrm{DM}%
}\right) _{\mathrm{thr}}$, but \emph{impedes} to form the GS, by making the
respective threshold, $\left( \omega _{\mathrm{DM}}\right) _{\mathrm{thr}}$,
higher. It is seen that for the DM the analytical approximation is
essentially more accurate than for the GS.
An example of a bound-state solution of linearized equations (\ref{U}) and (%
\ref{V}) of the DM type, numerically found at $\lambda =0.225$ and $\omega
=1.4$, i.e., \emph{below} the threshold value $\left( \omega _{\mathrm{DM}%
}\right) _{\mathrm{thr}}=1.5$, corresponding to the limit of $\lambda
\rightarrow 0$, is displayed in Fig. \ref{fig9}. The existence of the DM at
this point agrees with the right panel of Fig. \ref{fig8}.

\subsubsection{The analysis for large values of $\protect\lambda $}

In the opposite limit of large coupling constant $\lambda $, an analytical
approximation for the the discrete eigenvalues can be developed too. In this
case, $|\mu |$ may also be large, $\sim \lambda $. In the zero-order
approximation, one may neglect the derivative term in Eq.~(\ref{V}), to
obtain $V\approx -\left( \lambda /\mu \right) U$. Then, substituting this
relation back into the originally neglected derivative term, one obtains a
necessary correction to this relation:%
\begin{equation}
V\approx -\frac{\lambda }{\mu }U+\frac{\lambda }{2\mu ^{2}}\frac{d^{2}U}{%
dx^{2}}.  \label{VU}
\end{equation}%
The subsequent substitution of this expression in the linearized equation (%
\ref{U}) leads to an equation for $U$ which is tantamount to the usual
stationary linear Schr\"{o}dinger equation with the HO potential:%
\begin{equation}
\left( \mu -\frac{\lambda ^{2}}{\mu }+\omega \right) U+\frac{1}{2}\left( 1+%
\frac{\lambda ^{2}}{\mu ^{2}}\right) \frac{d^{2}U}{dx^{2}}-\frac{1}{2}%
x^{2}U=0.  \label{linear}
\end{equation}%
Then, the standard solution for the quantum-mechanical HO yields an equation
which determines the spectrum of the eigenvalues:%
\begin{equation}
\mu -\frac{\lambda ^{2}}{\mu }+\omega =\left( \frac{1}{2}+n\right) \sqrt{1+%
\frac{\lambda ^{2}}{\mu ^{2}}},  \label{quartic}
\end{equation}%
where $n=0,1,2,...$ is the quantum number. Taking into regard that $\lambda $
is now a large parameter, Eq.\ (\ref{quartic}) produces a final result for
the spectrum,%
\begin{equation}
\mu \approx -\lambda -\frac{\omega }{2}+\frac{1}{\sqrt{2}}\left( \frac{1}{2}%
+n\right) .  \label{muGS}
\end{equation}%
The spectrum remains equidistant in the current approximation, while further
corrections $\sim 1/\lambda $ give rise to terms $\sim \left( 1/2+n\right)
^{2}$, which break this property. As concerns the existence threshold for
the bound states, Eq.~(\ref{muGS}) predicts $\omega _{\mathrm{thr}}\approx
-2\lambda $. The coefficient in this relation is not accurate, as the
derivation is not valid for small $|\mu |$, but the implication is that, for
large $\lambda $, $\omega _{\mathrm{thr}}$ drops to negative values with a
large modulus, $\sim -\lambda $.

\begin{figure}[tbp]
\includegraphics[width=16cm]{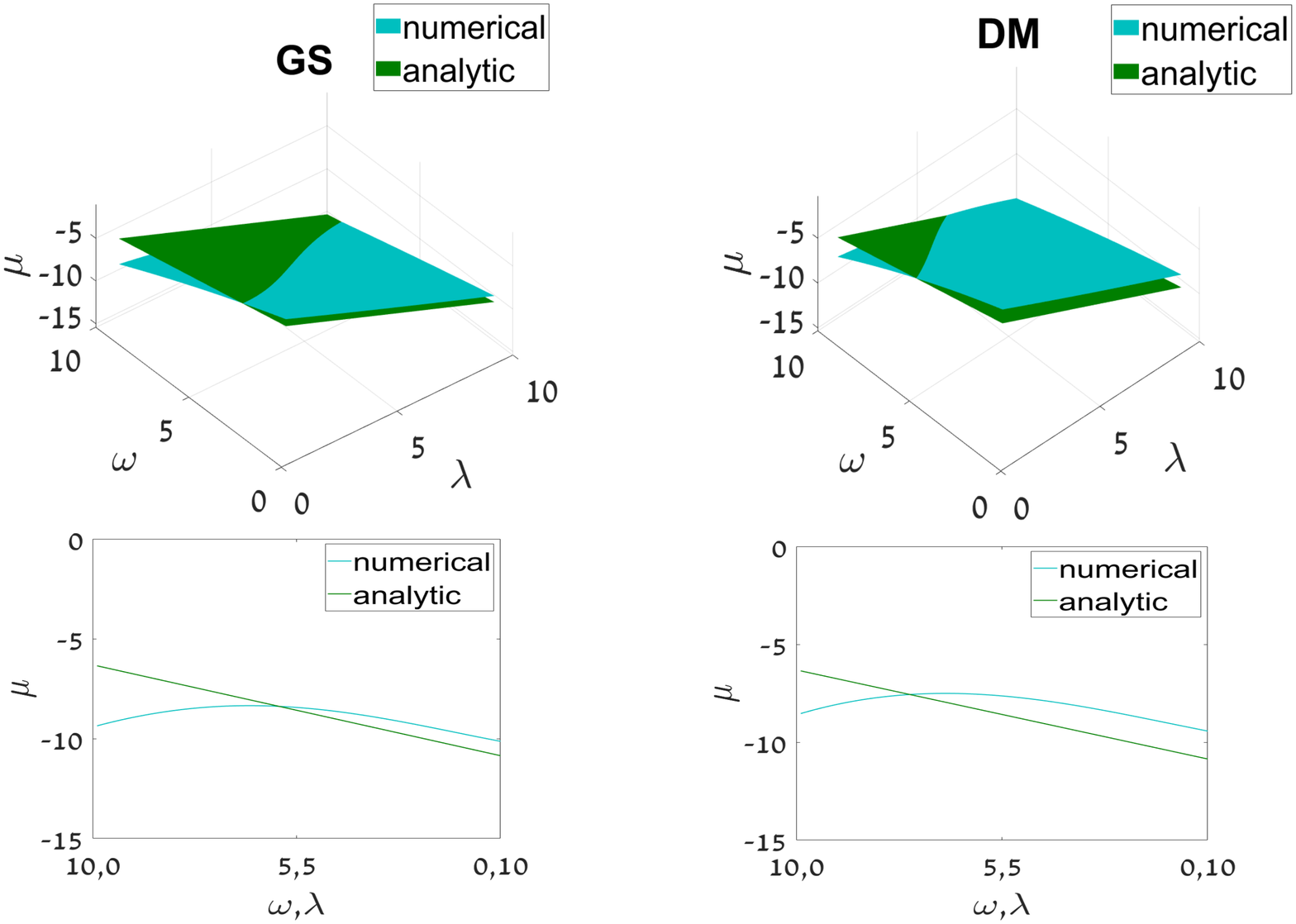}
\caption{{}The top row: left and right panels display the analytically
predicted eigenvalues given by Eq.~(\protect\ref{muGS}) with $n=0$ and $n=1$%
, for the ground state (GS) and dipole mode (DM), respectively, of the
half-trapped system, and their counterparts produced by the numerical
solution of linearized coupled equations (\protect\ref{U}) and (\protect\ref%
{V}), as functions of the linear-coupling constant, $\protect\lambda $, and
mismatch parameter, $\protect\omega $. The bottom row: cross sections of the
respective top panels along the diagonal connecting points $\left( \protect%
\lambda ,\protect\omega \right) =\left( 0.10\right) $ and $\left(
10,0\right) $. The results shown in these plots are relevant for relatively
large values of $\protect\lambda $.}
\label{fig10}
\end{figure}

The prediction of the GS and DM eigenvalues, given by Eq.~(\ref{muGS}) with $%
n=0$ and $n=1$, respectively, is compared to numerically found counterparts
in Fig. \ref{fig10}, which shows proximity between the analytical and
numerical results. The plots do not terminate in the displayed domain, i.e.,
they do not reach the existence boundary.

\subsubsection{Exact solutions for one- and two-dimensional \textit{bound
states in the continuum} (BIC) in the linear system}

A remarkable property of the coupled half-trapped system, represented by the
linearized version of Eqs. (\ref{asymm}) and (\ref{U}), (\ref{V}), is that
it admits particular spatially-confined solutions in an exact analytical
form. These are exceptional solutions, which, for an arbitrary value of the
linear-coupling constant, $\lambda $, exist at a single, specially selected,
value of the mismatch parameter, $\omega $, and with a single value of the
eigenvalue, $\mu $. First, it is possible to find an exact mode which is a
fundamental one (GS) in the $V$ component, and a \textit{second-order mode}
in $U$:%
\begin{gather}
U(x)=U_{0}\left[ \left( \lambda ^{2}-\frac{1}{2}\right) +x^{2}\right] \exp
\left( -\frac{x^{2}}{2}\right) ,  \notag \\
V(x)=-2\lambda U_{0}\exp \left( -\frac{x^{2}}{2}\right) ,  \notag \\
U_{0}^{2}=\pi ^{-1/2}\left( \lambda ^{4}+4\lambda ^{2}+1/2\right) ^{-1},
\notag \\
\omega =\frac{9}{4}-\frac{\lambda ^{2}}{2},~\mu =\frac{1}{2}\left( \lambda
^{2}+\frac{1}{2}\right) ,  \label{BIC1}
\end{gather}%
with amplitude $U_{0}$ defined by condition $P=1$, see Eq.~(\ref{P=1}). We
stress that, as seen in Eq.~(\ref{BIC1}), this exact solution may only have $%
\mu >0$, i.e., it is BIC (a \textit{bound state in the continuous spectrum}
\cite{BIC,BIC2,BIC3}), alias an \textit{embedded mode} \cite{embedded}. It
is worthy to note that this BIC mode and additional ones, presented below,
are found in the coupled system, with one component trapped in the HO\
potential.

In addition to the above spatially-even solution, an odd one of the BIC type
is available too. It is composed of a DM in the $V$ component and a \textit{%
third-order mode} in $U$. In the normalized form, i.e., with $P=1$ (as per
Eq.~(\ref{P=1})), the solution is%
\begin{gather}
U(x)=U_{0}x\left[ \left( \lambda ^{2}-\frac{3}{2}\right) +x^{2}\right] \exp
\left( -\frac{x^{2}}{2}\right) ,  \notag \\
V(x)=-2\lambda U_{0}x\exp \left( -\frac{x^{2}}{2}\right) ,  \notag \\
U_{0}^{2}=2\pi ^{-1/2}\left( \lambda ^{4}+4\lambda ^{2}+3/2\right) ^{-1},
\notag \\
\omega =\frac{11}{4}-\frac{\lambda ^{2}}{2},~\mu =\frac{1}{2}\left( \lambda
^{2}+\frac{3}{2}\right) ,  \label{BIC2}
\end{gather}%
which also exists at a single value of $\omega $, and with a single
eigenvalue, $\mu >0$. Note that both exact solutions, given by Eqs. (\ref%
{BIC1}) and (\ref{BIC2}), may exists at positive and negative values of $%
\omega $, as well as at $\omega =0$ (in Eqs. (\ref{BIC1}) and (\ref{BIC2}), $%
\omega =0$ at $\lambda ^{2}=9/2$ and $\lambda ^{2}=11/2$, respectively).

These exact solutions for BIC states in the two-component systems are
somewhat similar to those found in Ref.~\cite{SOC}, which addressed a system
of spin-orbit-coupled linear Gross-Pitaevskii equations for a binary BEC. In
that work, exact solutions were produced for a specially designed form of
the trapping potential.

\begin{figure}[tbp]
\includegraphics[width=16cm]{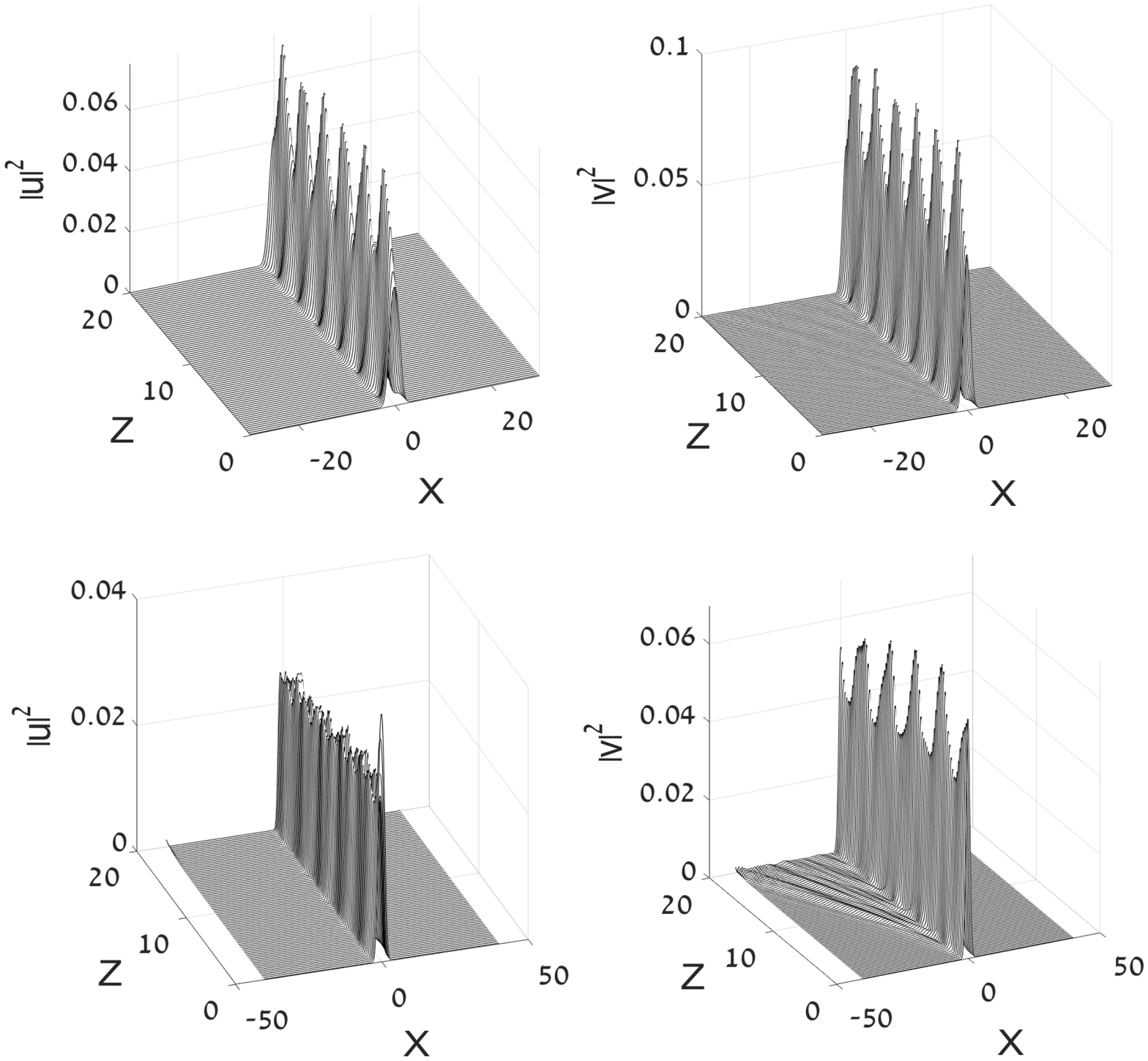}
\caption{The evolution initiated by the exceptional (BIC) exact solution (%
\protect\ref{BIC1}) of the system of linearized equations (\protect\ref{U})
and (\protect\ref{V}), as produced by simulations of the full nonlinear
half-trapped system (\protect\ref{asymm}), with the attractive SPM, $\protect%
\sigma =1$ in the top row, and repulsive $\protect\sigma =-1$ in the bottom
one. Other parameters are $g=0$, $\protect\lambda =7/2$ and$~\protect\omega %
=1/2$, which are related as per Eq.~(\protect\ref{BIC1}).}
\label{fig11}
\end{figure}

The exact solutions of the linearized system may be tried as inputs in
simulations of the full nonlinear system based on Eq.~(\ref{asymm}), with $%
\omega $ selected as per the solutions. The simulations, performed with
moderate values of the initial amplitude, produce a robust state with steady
internal oscillations, as shown in Fig. \ref{fig11} for the attractive ($%
\sigma =1$) and repulsive ($\sigma =-1$) SPM nonlinearity. In addition, the
simulations demonstrate weak emission of radiation in the untrapped ($v$)
component. In the case of the self-attraction (the left panel in Fig. \ref%
{fig7}), the radiation is almost invisible. The self-repulsion in the $v$
component, naturally, enhances the emission, which becomes visible in the
right panel of the figure. Still larger amplitudes of the input lead to
irregular oscillations and conspicuous emission of radiation (not shown here
in detail).

It may happen that the linearized system of Eqs. (\ref{U}) and (\ref{V})
produces isolated BIC/embedded modes in a numerical form at other values of
parameters. The present work does not aim to carry our comprehensive search
for such solutions. On the other hand, it is relevant to mention that a
straightforward two-dimensional\ (2D) extension of the present system
readily produces exceptional exact solutions for BIC/embedded modes of both
fundamental (GS, alias zero-vorticity) and vortex types.

The 2D extension of the linearized form of Eq.~(\ref{asymm}) is%
\begin{eqnarray}
iu_{z}+\frac{1}{2}\left( u_{xx}+u_{yy}\right) +\lambda v-\frac{1}{2}\left(
x^{2}+y^{2}\right) u &=&-\omega u,  \notag \\
&&  \label{2D} \\
iv_{z}+\frac{1}{2}\left( v_{xx}+v_{yy}\right) +\lambda u &=&0.  \notag
\end{eqnarray}%
While the realization of this model in optics in not straightforward, it may
be implemented for matter waves in a dual-core ``pancake-shaped" holder of
BEC \cite{Viskol,Arik}. In polar coordinates $\left( r,\theta \right) $,
particular exact solutions of linear equations (\ref{2D}), with all integer
values of the vorticity, $S=0,1,2,...$, are found as%
\begin{gather}
u=U_{0}r^{S}\left[ \frac{1}{2}\left( \lambda ^{2}-1-S\right) +r^{2}\right]
\exp \left( -i\mu z+iS\theta -\frac{r^{2}}{2}\right) ,  \notag \\
v=-2\lambda U_{0}r^{S}\exp \left( -i\mu z+iS\theta -\frac{r^{2}}{2}\right) ,
\notag \\
\omega =\frac{1}{2}\left( 5+S-\lambda ^{2}\right) ,~\mu =\frac{1}{2}\left(
\lambda ^{2}+1+S\right) ,  \label{vort}
\end{gather}%
with arbitrary amplitude $U_{0}$ (the 2D solution of the GS type corresponds
to $S=0$ in Eq.~(\ref{vort})). As well as in 1D solutions (\ref{BIC1}) and (%
\ref{BIC2}), $\mu $ takes only positive values in the 2D solution, hence it
also represents states of the BIC/embedded type.

\subsection{The nonlinear half-trapped system}

\subsubsection{The Thomas-Fermi approximation (TFA)}

In the presence of the self-defocusing nonlinearity, $\sigma =-1$ in Eqs. (%
\ref{U}) and (\ref{V}), it is relevant to apply TFA to finding the GS of the
half-trapped system, omitting the second derivatives in both equations \cite%
{Pit}, and keeping condition $\mu <0$, which is necessary for the existence
of a generic (non-BIC) localized state in the $V$ component. Then, Eq.~(\ref%
{V}) with $g=0$ (the XPM coupling is omitted here) is solved as%
\begin{equation}
U=\left( V/\lambda \right) \left( V^{2}-\mu \right) ,  \label{UTFA}
\end{equation}%
and Eq.~(\ref{U}) amounts to an algebraic equation for the squared amplitude
$W\equiv -V^{2}/\mu >0$, \textit{viz}.,%
\begin{equation}
mW\left( W+1\right) ^{3}+\xi ^{2}W=\xi _{0}^{2}-\xi ^{2},  \label{W}
\end{equation}%
where%
\begin{eqnarray}
m &\equiv &-\frac{2\mu }{\lambda ^{2}}>0,~\xi \equiv -\frac{x}{\mu },
\label{mxi} \\
\xi _{0}^{2} &\equiv &-\frac{2}{\mu ^{3}}\left[ \lambda ^{2}-\mu \left(
\omega +\mu \right) \right] ,~  \label{xi0}
\end{eqnarray}%
and the applicability condition for TFA is easily shown to be $m\ll 1$. The
TFA solution exists under condition $\xi _{0}^{2}>0$ (see Eq.~(\ref{xi0})),
i.e., in a finite interval (\textit{bandgap}) of values of the propagation
constant,%
\begin{equation}
0<-\mu <\sqrt{\lambda ^{2}+\omega ^{2}/4}+\omega /2\equiv -\mu _{0}.
\label{bandgap}
\end{equation}

\begin{figure}[tbp]
\includegraphics[width=12cm]{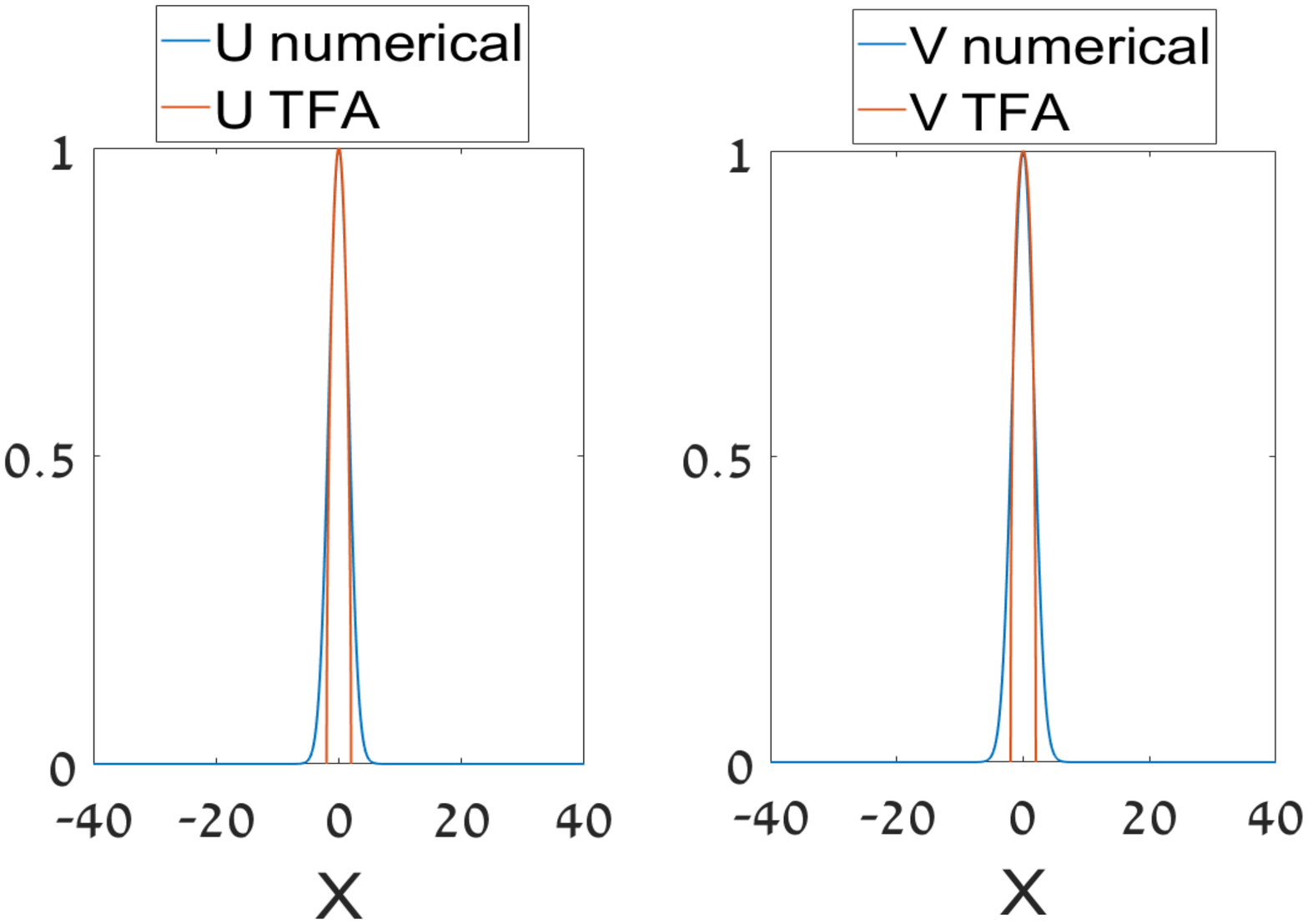}
\caption{{}A typical example of the GS solution predicted by TFA
(Thomas-Fermi approximation) for the half-trapped system, as per Eqs. (%
\protect\ref{UTFA})-(\protect\ref{xi0}), for $\protect\sigma =-1$, $g=0$, $%
\protect\lambda =8$, $\protect\mu =-4$, and its comparison to the
numerically found counterpart. The respective value of parameter $m$ (see
Eq.~(\protect\ref{mxi})), which should be small for the applicability of
TFA, is $m=0.125$.}
\label{fig12}
\end{figure}

Outside of the bandgap, i.e., at $\mu <$ $\mu _{0}<0$, the TFA solution does
not exist. In the bandgap, Eq.~(\ref{W}) produces a usual GS profile, with a
single value of $W$ corresponding to each $\xi ^{2}$ from region $\xi
^{2}<\xi _{0}^{2}$ (see Fig. \ref{fig12}, which displays a typical example
of the TFA-predicted GS and its comparison with the numerically found
counterpart). The solution vanishes at the border points, $\xi =\pm \xi _{0}$%
, and is equal to zero at $\xi ^{2}>\xi _{0}^{2}$, so that the derivative of
the TFA solution, $dW/d\xi $, is discontinuous at the border points, which
is a usual peculiarity of the TFA \cite{Fetter,Pit}.

Note that, in the limit of large $\lambda $, the bandgap's width, as given
by Eq.~(\ref{bandgap}), is $-\mu _{0}\approx \lambda +\omega /2$, which is
close to the largest value of $-\mu $ predicted by Eq.~(\ref{muGS}) for the
GS ($n=0$) in the linearized half-trapped system. Finally, the TFA solution
may be cast in a simple explicit form close to the edge of the bandgap,
i.e., at%
\begin{eqnarray}
0 &<&\delta \mu \equiv \mu -\mu _{0}\ll -\mu _{0},  \label{edge} \\
\xi _{0}^{2} &\approx &-\left( 4/\mu _{0}^{3}\right) \sqrt{\lambda
^{2}+\omega ^{2}/4}\delta \mu  \label{edge2}
\end{eqnarray}%
In this case, Eqs. (\ref{W}) and (\ref{UTFA}) simplify to%
\begin{equation}
U^{2}\approx \left\{
\begin{array}{c}
\left( \mu _{0}^{2}/2\right) \left( \xi _{0}^{2}-\xi ^{2}\right) ,~\mathrm{%
at~}\xi ^{2}<\xi _{0}^{2}, \\
0~\mathrm{at~}\xi ^{2}>\xi _{0}^{2},%
\end{array}%
\right.  \label{U^2}
\end{equation}%
\begin{equation}
V^{2}\approx \left\{
\begin{array}{c}
\left( \lambda ^{2}/2\right) \left( \xi _{0}^{2}-\xi ^{2}\right) ,~\mathrm{%
at~}\xi ^{2}<\xi _{0}^{2}, \\
0~\mathrm{at~}\xi ^{2}>\xi _{0}^{2}.%
\end{array}%
\right.  \label{V^2}
\end{equation}

Expressions (\ref{edge2})-(\ref{V^2}) make it possible to calculate the
total power of the GS (see Eq.~(\ref{P})),%
\begin{equation}
P\approx \frac{16}{3}\left( \lambda ^{2}+\mu _{0}^{2}\right) \left( \lambda
^{2}+\frac{\omega ^{2}}{4}\right) ^{3/4}\left( -\mu _{0}\right)
^{-7/2}\left( \delta \mu \right) ^{3/2}.  \label{Pmu}
\end{equation}%
Note that relation (\ref{Pmu}) satisfies the \textit{anti-Vakhitov-Kolokolov
criterion}, $dP/d\left( \delta \mu \right) >0$, which is a necessary
condition for stability of localized states supported by self-repulsive
nonlinearity \cite{anti}, as is the case in the present setting (the
Vakhitov-Kolokolov criterion proper, $dP/d\left( \delta \mu \right) <0$, is
a well-known necessary condition for stability of states in models with
self-attraction \cite{VK,Berge,Fibich}).

\subsubsection{Existence boundaries for nonlinear states}

\begin{figure}[tbp]
\includegraphics[width=11.0cm]{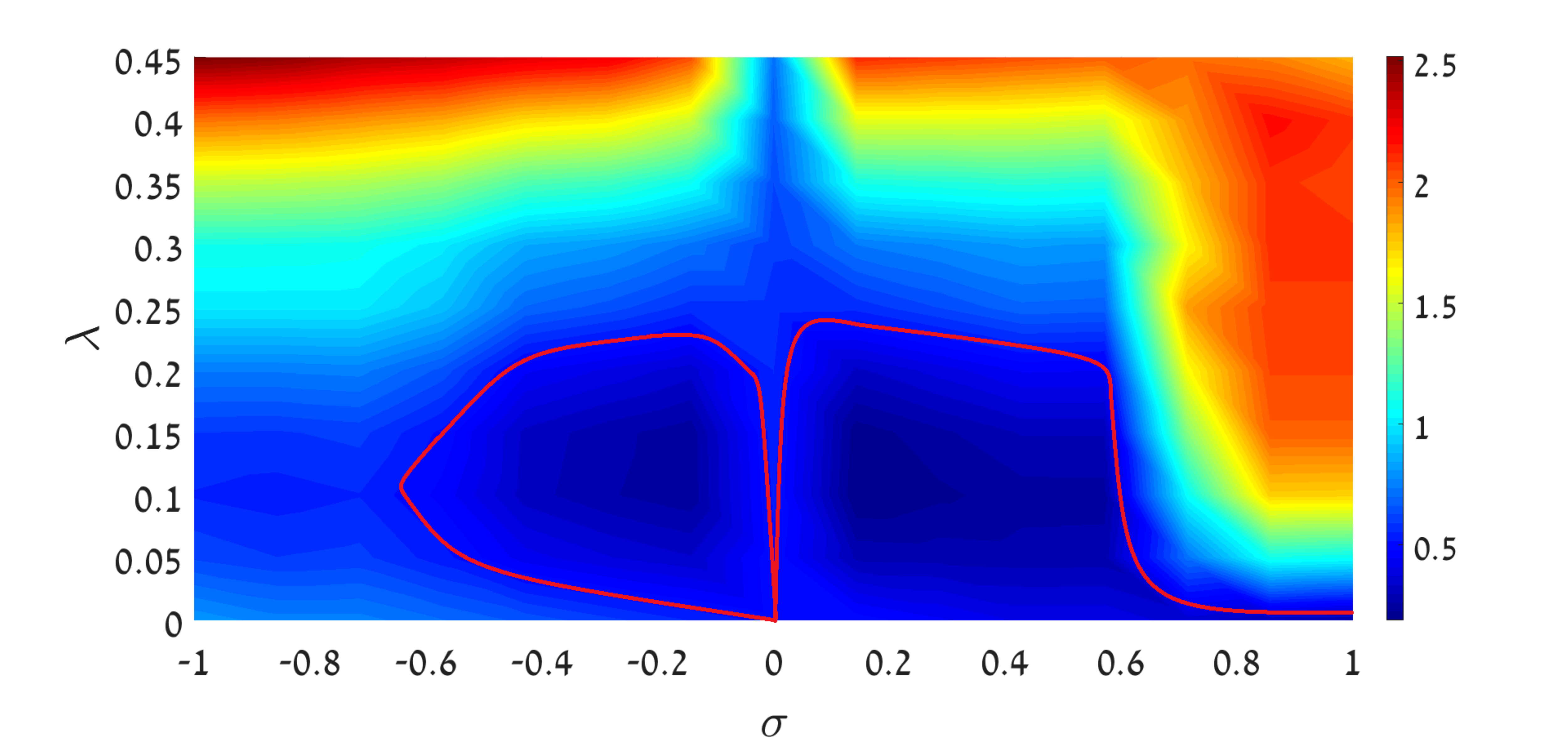}
\caption{{}The heatmap of threshold values of the mismatch parameter, $%
\left( \protect\omega _{\mathrm{GS}}\right) _{\mathrm{thr}}$, in the
half-trapped system, based on Eqs. (\protect\ref{U}) and (\protect\ref{V}).
For given values of the nonlinearity and linear-coupling coefficients, $%
\protect\sigma $ and $\protect\lambda $, the stable GS (ground state),
subject to the normalization condition $P=1$ (see Eq.~(\protect\ref{P=1})),
exist above the threshold, i.e., at $\protect\omega \geq \left( \protect%
\omega _{\mathrm{GS}}\right) _{\mathrm{thr}}$. Positive and negative values
of $\protect\sigma $ correspond to the attractive and repulsive sign of the
self-interaction, respectively. The nontrivial region is one confined by red
lines, in which the result is $\left( \protect\omega _{\mathrm{GS}}\right) _{%
\mathrm{thr}}<1/2$, i.e., the nonlinearity and linear coupling help to
maintain self-trapped GSs in the parameter area where the decoupled system,
with $\protect\lambda =0$, cannot create such states. }
\label{fig13}
\end{figure}

\begin{figure}[tbp]
\includegraphics[width=16cm]{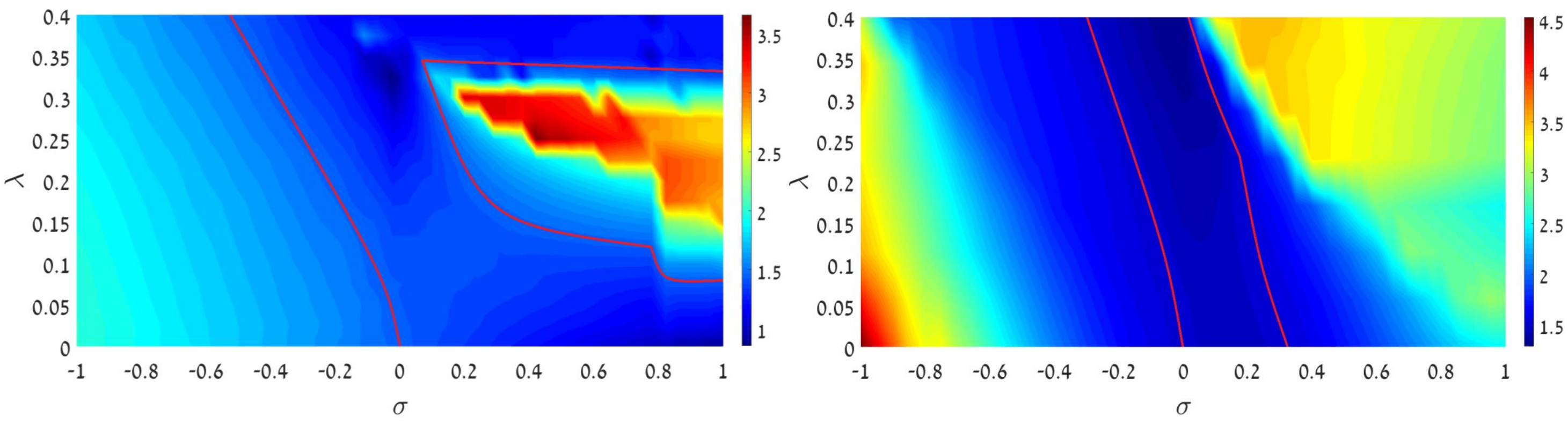}
\caption{{}The same as in Fig. \protect\ref{fig13}, but for DMs (dipole
modes), obtained as numerical solutions to Eqs. (\protect\ref{U}) and (%
\protect\ref{V}) with $g=0$ and $g=1$ (the left and right panels,
respectively). In this case, the nontrivial region, located between the red
boundaries, is one with $\left( \protect\omega _{\mathrm{DM}}\right) _{%
\mathrm{thr}}<3/2$.}
\label{fig14}
\end{figure}

The shrinkage of the existence region for GS solutions, and its expansion
for DM ones, in the linear version of the coupled half-trapped system, shown
in Fig. \ref{fig8}, suggests to identify existence boundaries of the same
states in the full nonlinear system. Numerical data, necessary for the
delineation of the existence region of the GSs and DMs in the nonlinear
system, were collected by solving Eqs. (\ref{U}) and (\ref{V}) for spatially
even and odd modes with the fixed total power, $P=1$ (as per Eq.~(\ref{P=1}%
)), while the linear-coupling coefficient, $\lambda $, and the nonlinearity
coefficient, $\sigma $, were varied, the latter one taking both positive and
negative values (for the self-attraction/repulsion). The results are
summarized by the heatmap in Figs. \ref{fig13} (for the GS) and \ref{fig14}
(for the DM), which show threshold values of the mismatch parameter, defined
as per Eq.~(\ref{thr}).

Nontrivial parametric areas for the GS and DM solutions are identified as
domains with, respectively, $\left( \omega _{\mathrm{GS}}\right) _{\mathrm{%
thr}}<1/2$ and $\left( \omega _{\mathrm{DM}}\right) _{\mathrm{thr}}<3/2$.
The former one is surrounded by red lines in Fig. \ref{fig13}, and its
counterpart for the DMs is located between red lines in Fig. \ref{fig14}. In
these areas, the coupled half-trapped system maintains stable localized GSs
at $\omega <1/2$, and DMs at $\omega <3/2$, while in the absence of the
coupling they may only exist at $\omega >1/2$ and $\omega >3/2$,
respectively.

Note that the vertical cross-section of the heatmap in Fig. \ref{fig13},
drawn through $\sigma =0$ (along which the system remains linear), that
starts from $\lambda =0$, belongs to the area with $\left( \omega _{\mathrm{%
GS}}\right) _{\mathrm{thr}}>1/2$, in agreement with the result for the
linear system, which shows that the coupling impedes the existence of the
GS, see Fig. \ref{fig8}(a) and Eq.~(\ref{smalllambda}). Further, Fig. \ref%
{fig13} demonstrates that the SPM\ nonlinearity of either sign facilitates
the creation of GS in parameter regions surrounded by red lines. Naturally,
the self-attraction ($\sigma >0$) helps to create such states starting from
arbitrarily small values of $\lambda $, while the self-repulsion ($\sigma <0$%
) is able to do it in the region separated by a gap from $\lambda =0$.
Nevertheless, at $\lambda >0.24$ the detrimental effect of the linear
coupling cannot be outweighed by the SPM nonlinearity.

In Fig. \ref{fig14}, the vertical cross section corresponding to the linear
system ($\sigma =0$) entirely belongs to the nontrivial area, with $\left(
\omega _{\mathrm{DM}}\right) _{\mathrm{thr}}<3/2$, also in agreement with
Eq.~(\ref{smalllambda}) and Fig. \ref{fig8}(b). Figure \ref{fig14}
demonstrates that the nonlinearity, generally, impedes the maintenance of
the localized DMs. This conclusion is supported, in particular, by the
comparison of panels (a) and (b) in the figure, which shows that the
addition of the attractive XPM nonlinearity with the to SPM conspicuously
reduces the remaining nontrivial area.

\section{Conclusion}

The objective of this work is to analyze new effects in the symmetric and
asymmetric systems of linearly coupled fields, which are subject to the
action of the HO (harmonic-oscillator) trapping potential and cubic
self-attraction or repulsion. The system can be implemented in nonlinear
optics and BEC. In the symmetric system, with identical HO potentials
applied to both components, we focus on the consideration of JO (Josephson
oscillations) in the system, by launching, in one component, an input in the
form of the GS (ground state) or DM (dipole mode) of the HO potential. On
the basis of systematically collected numerical data, we have identified two
transitions in the system's dynamics, which occur with the increase of the
input's power in the case of the self-attraction. The first is SSB
(spontaneous symmetry breaking) between the linearly coupled components in
the dynamical JO state. At a higher power, the nonlinearity causes a
transition from regular JO, initiated by the GS input, to chaotic dynamics.
This transition is identified through consideration of spectral
characteristics of the dynamical regime. The input in the form of the DM
undergoes the chaotization at essentially smaller powers than the dynamical
regime initiated by the GS input, which is followed by the SSB at slightly
higher powers. In the case of self-repulsion, SSB does not occur, while the
chaotization takes place in a weak form, in a small part of the parameter
space.

In the half-trapped system, with the HO potential acting on a single
component, a nontrivial issue is identification of the system's linear
spectrum, i.e., a parameter region in which the linearized system maintains
trapped binary (two-component) modes. This problem is solved here
analytically in the limit cases of weak and strong linear coupling, and in
the numerical form in the general case. In particular, the linear coupling
between the components leads to the shrinkage of the spectral band in which
the GS exists, and expansion of the existence band for the DM. The existence
region for trapped states in the full nonlinear system is identified
numerically, and such states are constructed analytically by means of the
TFA (Thomas-Fermi approximation). In addition, exceptional solutions of the
linearized system of the BIC (bound-state-in-continuum), alias embedded,
type were found in the exact analytical form, in both the 1D and 2D
settings, the 2D solution being found with an arbitrary value of the
vorticity.

The work may be extended by considering inputs in the form of higher-order
HO eigenstates. Another relevant direction for the extension of the analysis
is a systematic study of the 2D system.

\section*{Acknowledgment}

This work was supported, in part, by the Israel Science Foundation through
grant No. 1286/17.


\begin{thebibliography}{99}
\bibitem{Pet} C. J. Pethick and H. Smith, \textit{Bose-Einstein Condensation
in Dilute Gases} (Cambridge University Press, Cambridge, 2002).

\bibitem{Pit} L. P. Pitaevskii and S. Stringari, \textit{Bose-Einstein
Condensation} (Oxford University Press, Oxford 2003).

\bibitem{Pan} P. G. Kevrekidis, D. J. Frantzeskakis, and R. Carretero-Gonz%
\'{a}lez, \textit{Emergent Nonlinear Phenomena in Bose-Einstein Condensates:
Theory and Experiment} (Springer: Heidelberg, 2008).

\bibitem{Schneider} B. I. Schneider and D. L. Feder, Numerical approach to
the ground and excited states of a Bose-Einstein condensed gas confined in a
completely anisotropic trap, Phys. Rev. A \textbf{59}, 2232-2242 (1999).

\bibitem{Sadhan} S. K. Adhikari, Numerical solution of the two-dimensional
Gross-Pitaevskii equation for trapped interacting atoms, Phys. Lett. A
\textbf{265}, 91-96 (2000).

\bibitem{Busch} T. Busch and J. R. Anglin, Motion of dark solitons in
trapped Bose-Einstein condensates, Phys. Rev. Lett. \textbf{84}, 2298-2301
(2000).

\bibitem{Turitsyn} Y. S. Kivshar, T. J. Alexander, and S. K. Turitsyn,
Nonlinear modes of a macroscopic quantum oscillator, Phys. Lett. \textbf{278}%
, 225-230 (2001).

\bibitem{Alexander} T. J. Alexander and L. Berg\'{e}, Ground states and
vortices of matter-wave condensates and optical guided waves, Phys. Rev. E
\textbf{65}, 026611 (2002).

\bibitem{Huang} G. Huang, J. Szeftel and S. Zhu, Dynamics of dark solitons
in quasi-one-dimensional Bose-Einstein condensates, Phys. Rev. A \textbf{65}%
, 053605 (2002).

\bibitem{Nick} N. G. Parker, N. P. Proukakis, M. Leadbeater, and C. S.
Adams, Soliton-sound interactions in quasi-one-dimensional Bose-Einstein
condensates, Phys. Rev. Lett. \textbf{90}, 220401 (2003).

\bibitem{Pelinovsky} D. E. Pelinovsky, D. J. Frantzeskakis, and P. G.
Kevrekidis, Oscillations of dark solitons in trapped Bose-Einstein
condensates, Phys. Rev. E \textbf{72}, 016615 (2005).

\bibitem{Brazhnyi} V. A. Brazhnyi and V. V. Konotop, Stable and unstable
vector dark solitons of coupled nonlinear Schr\"{o}dinger equations:
Application to two-component Bose-Einstein condensates, Phys. Rev. E \textbf{%
72}, 026616 (2005).

\bibitem{Nick2} N. G. Parker, N. G. Proukakis, and C. S. Adams, Dark soliton
decay due to trap anharmonicity in atomic Bose-Einstein condensates, Phys.
Rev. A \textbf{81}, 033606 (2010).

\bibitem{Newcastle} T. Bland, N. G. Parker, N. P. Proukakis, and B. A.
Malomed, Probing quasi-integrability of the Gross-Pitaevskii equation in a
harmonic-oscillator potential, J. Phys. B: At. Mol. Opt. Phys. \textbf{51},
205303 (2018).

\bibitem{Merhasin} M. I. Merhasin, B. A. Malomed, and R. Driben, Transition
to miscibility in a binary Bose-Einstein condensate induced by linear
coupling, J. Physics B \textbf{38}, 877-892 (2005).

\bibitem{bright-dark} H. E. Nistazakis, D. J. Frantzeskakis, P. G.
Kevrekidis, B. A. Malomed, and R. Carretero-Gonzalez, Bright-dark soliton
complexes in spinor Bose-Einstein condensates, Phys. Rev. A \textbf{77},
033612 (2008).

\bibitem{Liu} D.-S. Wang, X.-H. Hu, J. Hu, and W. M. Liu, Quantized
quasi-two-dimensional Bose-Einstein condensates with spatially modulated
nonlinearity, Phys. Rev. A \textbf{81}, 025604 (2010).

\bibitem{Viskol} Z. Chen, Y. Li, B. A. Malomed, and L. Salasnich,
Spontaneous symmetry breaking of fundamental states, vortices, and dipoles
in two and one-dimensional linearly coupled traps with cubic
self-attraction, Phys. Rev. A \textbf{96}, 033621 (2016).

\bibitem{RMP} Y. S. Kivshar and B. A, Malomed, Dynamics of solitons in
nearly integrable systems, Rev. Mod. Phys. \textbf{61}, 763-915 (1989).

\bibitem{Agrawal} S. Raghavan and G. P. Agrawal, Spatiotemporal solitons in
inhomogeneous nonlinear media, Opt. Commun. \textbf{180}, 377-382 (2000).

\bibitem{Zezyu} D. A. Zezyulin, G. L. Alfimov, and V. V. Konotop, Nonlinear
modes in a complex parabolic potential, Phys. Rev. A \textbf{81}, 013606
(2010).

\bibitem{Stathis} E. G. Charalampidis, P. G. Kevrekidis, D. J.
Frantzeskakis, and B. A. Malomed, Dark-bright solitons in coupled NLS
equations with unequal dispersion coefficients, Phys. Rev. E \textbf{91},
012924 (2015).

\bibitem{Thaw} T. Mayteevarunyoo, B. A. Malomed, and D. V. Skryabin, One-
and two-dimensional modes in the complex Ginzburg-Landau equation with a
trapping potential, Opt. Exp. \textbf{26}, 8849-8865 (2018).

\bibitem{Josephson network} M. Leib, F. Deppe, A. Marx, R. Gross, and M. J.
Hartmann, Networks of nonlinear superconducting transmission line
resonators, New J. Phys. \textbf{14}, 075024 (2012).

\bibitem{KA} Y. S. Kivshar and G. P. Agrawal, \textit{Optical Solitons: From
Fibers to Photonic Crystals }(Academic Press: San Diego, 2003).

\bibitem{Peyrard} T. Dauxois, M. Peyrard, \textit{Physics of Solitons}
(Cambridge University Press: Cambridge, 2006).

\bibitem{Morsch} O. Morsch and M. Oberthaler, Dynamics of Bose-Einstein
condensates in optical lattices, Rev. Modern Phys. \textbf{78}, 179-215
(2006).

\bibitem{PhCr1} J. D. Joannopoulos, S. G. Johnson, J. N. Winn, and R. D.
Meade, \textit{Photonic Crystals: Molding the Flow of Light} (Princeton
University Press: Princeton, 2008).

\bibitem{PhCr2} M. Skorobogatiy, and J. Yang, Fundamentals of Photonic
Crystal Guiding (Cambridge University Press: Cambridge, 2008).

\bibitem{PhCr3} E. A. Cerda-Mendez, D. Sarkar, D. N. Krizhanovskii, S. S.
Gavrilov, K. Biermann, M. S. Skolnick, P. V. Santos, Exciton-polariton gap
solitons in two-dimensional lattices, Phys. Rev. Lett. \textbf{111}, 146401
(2013).

\bibitem{KB} V. A. Brazhnyi and V. V. Konotop, Theory of nonlinear matter
waves in optical lattices, Modern Phys. Lett. B \textbf{18}, 627-651 (2004).

\bibitem{HS} H. Sakaguchi and B. A. Malomed, Dynamics of positive- and
negative-mass solitons in optical lattices and inverted traps, J. Phys. B
\textbf{37}, 1443-1459 (2004).

\bibitem{Zakharov} V. E. Zakharov, S. V. Manakov, S. P. Novikov, and L. P.
Pitaevskii, \textit{Theory of Solitons: Inverse Scattering Method} (Nauka:
Moscow, 1980; English translation: Consultants Bureau, New York, 1984).

\bibitem{turbulence} V. Zakharov, F. Dias, and A. Pushkarev, One-dimensional
wave turbulence, Phys. Rep. \textbf{398}, 1-65 (2004).

\bibitem{Mazets} I. E. Mazets and J. Schmiedmayer, Thermalization in a
quasi-one-dimensional ultracold bosonic gas, New J. Phys. \textbf{12},
055023 (2010).

\bibitem{Nick1} S. P. Cockburn, A. Negretti, N. P. Proukakis, and C. Henkel,
Comparison between microscopic methods for finite-temperature Bose gases,
Phys. Rev. A \textbf{83}, 043619 (2011).

\bibitem{Mazets2} P. Grisins and I. E. Mazets, Thermalization in a
one-dimensional integrable system, Phys. Rev. A \textbf{84}, 053635 (2011).

\bibitem{Kheruntsyan} K. F. Thomas, M. J. Davis, and K. V. Kheruntsyan,
Thermalization of a quantum Newton's cradle in a one-dimensional
quasicondensate, Phys. Rev. A \textbf{103}, 023315 (2021).

\bibitem{twist} M. Decker, M. Ruther, C. E, Kriegler, J. Zhou, C. M.
Soukoulis, S. Linden, and M. Wegener, Strong optical activity from
twisted-cross photonic metamaterials, Opt. Lett. \textbf{34}, 2501-2503
(2009).

\bibitem{GHz1} R. J. Ballagh, K. Burnett, and T. F. Scott, Theory of an
output coupler for Bose-Einstein condensed atoms, Phys. Rev. Lett. \textbf{78%
}, 1607-1611 (1997).

\bibitem{GHz2} P. \"{O}hberg and S. Stenholm, Internal Josephson effect in
trapped double condensates, Phys. Rev. A \textbf{59}, 3890-3895 (1999).

\bibitem{GHz3} D. T. Son and M. A. Stephanov, Domain walls of relative phase
in two-component Bose-Einstein condensates, Phys. Rev. A \textbf{65}, 063621
(2002).

\bibitem{GHz4} S. D. Jenkins and T. A. B. Kennedy, Dynamic stability of
dressed condensate mixtures, Phys. Rev. A \textbf{68}, 053607 (2003).

\bibitem{book} B. A. Malomed, editor: \textit{Spontaneous Symmetry Breaking,
Self-Trapping, and Josephson Oscillations}, (\noindent Springer-Verlag:
Berlin and Heidelberg, 2013).

\bibitem{Ignac} N. V. Hung, L. X. T. Tai, I. Bugar, M. Longobucco, R. Buzcy%
\'{n}ski, B. A. Malomed, and M. Trippenbach, Reversible ultrafast soliton
switching in dual-core highly nonlinear optical fibers, Opt. Lett. \textbf{45%
}, 5221-5224 (2020).

\bibitem{Mineev} V. P. Mineev, The theory of the solution of two near-ideal
Bose gases, Zh. Eksp. Teor. Fiz. \textbf{67}, 263-272 (1974) [English
translation: Sov. Phys. - JETP \textbf{40}, 132-136 (1974)].

\bibitem{OptJO6} B. A. Malomed, Solitons and nonlinear dynamics in dual-core
optical fibers, In: \textit{Handbook of Optical Fibers} (G.-D. Peng, Editor:
Springer, 2018), pp. 421-474.

\bibitem{ST} M. V. Tratnik and J. E. Sipe, Bound solitary waves in a
birefringent optical fiber, Phys. Rev. A \textbf{38}, 2011-2017 (1988).

\bibitem{BECJO1} G. J. Milburn, J. Corney, E. M. Wright, and D. F. Walls,
Quantum dynamics of an atomic Bose-Einstein condensate in a double-well
potential, Phys. Rev. A \textbf{55}, 4318 (1997).

\bibitem{BECJO2} A. Smerzi, S. Fantoni, S. Giovanazzi, and S. R. Shenoy,
Quantum coherent atomic tunneling between two trapped Bose-Einstein
condensates, Phys. Rev. Lett. \textbf{79}, 4950 (1997).

\bibitem{BECJO3} M. Albiez, R. Gati, J. F\"{o}lling, S. Hunsmann, M.
Cristiani, and M. K. Oberthaler, Direct observation of tunneling and
nonlinear self-trapping in a single bosonic Josephson junction, Phys. Rev.
Lett. \textbf{95}, 010402 (2005).

\bibitem{BECJO4} Y. Shin, G.-B. Jo, M. Saba, T. A. Pasquini, W. Ketterle,
and D. E. Pritchard, Optical weak link between two spatially separated
Bose-Einstein condensates, Phys. Rev. Lett. \textbf{95}, 170402 (2005).

\bibitem{BECJO5} S. Levy, E. Lahoud, I. Shomroni, and J. Steinhauer, The
a.c. and d.c. Josephson effects in a Bose--Einstein condensate, Nature
\textbf{449}, 579-583 (2007).

\bibitem{BECJO6} Z. Chen, Y. Li, and B. A. Malomed, Josephson oscillations
of chirality and identity in two-dimensional solitons in spin-orbit-coupled
condensates, Phys. Rev. Research \textbf{2}, 033214 (2020).

\bibitem{HS2} H. Sakaguchi and B. A. Malomed, Symmetry breaking in a
two-component system with repulsive interactions and linear coupling, Comm.
Nonlin. Sci. Num. Sim. \textbf{92}, 105496 (2020).

\bibitem{OptJO1} C. Par\'{e} and M. F\l orja\'{n}czyk, Approximate model of
soliton dynamics in all-optical couplers, Phys. Rev. A \textbf{41}, 6287
(1990).

\bibitem{OptJO2} A. I. Maimistov, Propagation of a light pulse in nonlinear
tunnel-coupled optical waveguides, Kvant. Elektron. \textbf{18}, 758 (1991)
[English translation: Sov. J. Quantum Electron. \textbf{21}, 687 (1991)].

\bibitem{OptJO3} I. M. Uzunov, R. Muschall, M. Goelles, Y. S. Kivshar, B. A.
Malomed, and F. Lederer, Pulse switching in nonlinear fiber directional
couplers, Phys. Rev. E \textbf{51}, 2527-2536 (1995).

\bibitem{OptJO4} M. Abbarchi, A. Amo, V. G. Sala, D. D. Solnyshkov, H.
Flayac, L. Ferrier, I. Sagnes, E. Galopin, A. Lema\^{\i}tre, G. Malpuech,
and J. Bloch, Macroscopic quantum self-trapping and Josephson oscillations
of exciton polaritons, Nature Phys. \textbf{9}, 275 (2013).

\bibitem{Osgood} N.-C. Panoiu, B. A. Malomed, and R. M. Osgood, Jr.,
Semidiscrete solitons in arrayed waveguide structures with Kerr
nonlinearity, Phys. Rev. A \textbf{78}, 013801 (2008).

\bibitem{BIC} F. H. Stillinger and D. R. Herrick, Bound states in continuum,
Phys. Rev. A \textbf{11}, 446-454 (1975).

\bibitem{BIC2} A. Kodigala, T. Lepetit, Q. Gu, B. Bahari, Y. Fainman, and B.
Kante, Lasing action from photonic bound states in continuum, Nature \textbf{%
54}1, 196-199 (2017).

\bibitem{BIC3} B. Midya and V. V. Konotop, Coherent-perfect-absorber and
laser for bound states in a continuum, Opt. Lett. 43, 607-610 (2018).

\bibitem{embedded} A. R. Champneys, B. A. Malomed, J. Yang, and D. J. Kaup,
\textquotedblleft Embedded solitons": solitary waves in resonance with the
linear spectrum, Physica D \textbf{152-153}, 340-354 (2001).

\bibitem{Oberthaler} M. Albiez, R. Gati, J. Folling, S. Hunsmann, M.
Cristiani, and M. K. Oberthaler, Direct observation of tunneling and
nonlinear self-trapping in a single bosonic Josephson junction, Phys. Rev.
Leyy. 95, 010402 (2005).

\bibitem{double-well} D. Ananikian and T. Bergeman, Gross-Pitaevskii
equation for Bose particles in a double-well potential: Two-mode models and
beyond, Phys. Rev. A \textbf{73}, 013604 (2006).

\bibitem{Konotop} Y. V. Kartashov, V. V. Konotop, and V. A. Vysloukh,
Dynamical suppression of tunneling and spin switching of a
spin-orbit-coupled atom in a double-well trap, Phys. Rev. A \textbf{97},
063609 (2018).

\bibitem{Iooss} G. Iooss and D. D. Joseph, \textit{Elementary Stability
Bifurcation Theory} (Springer, New York, 1980).

\bibitem{LL} L. D. Landau and E. M. Lifshitz, \textit{Quantum Mechanics}
(Nauka Publishers, Moscow, 1974).

\bibitem{SOC} Y. V. Kartashov, V. V. Konotop, and L. Torner1, Bound states
in the continuum in spin-orbit-coupled atomic systems, Phys. Rev. A \textbf{%
96}, 033619 (2017).

\bibitem{Arik} M. C. P. dos Santos, B. A. Malomed, and W. B. Cardoso,
Double-layer Bose-Einstein condensates: A quantum phase transition in the
transverse direction, and reduction to two dimensions, Phys. Rev. E 102,
042209 (2020).

\bibitem{Fetter} A. L. Fetter and D. L. Feder, Beyond the Thomas-Fermi
approximation for a trapped condensed Bose-Einstein gas, Phys. Rev. A
\textbf{58}, 3185-3194 (1998).

\bibitem{anti} H. Sakaguchi and B. A. Malomed, Solitons in combined linear
and nonlinear lattice potentials, Phys. Rev. A \textbf{81}, 013624 (2010).

\bibitem{VK} N. G. Vakhitov and A. A. Kolokolov, Stationary solutions of the
wave equation in a medium with nonlinearity saturation, Radiophys. Quant.
Electron. \textbf{16}, 783-789 (1973).

\bibitem{Berge} L. Berg\'{e}, Wave collapse in physics: principles and
applications to light and plasma waves, Phys. Rep. \textbf{303}, 259-370
(1998).

\bibitem{Fibich} G. Fibich, \textit{The Nonlinear Schr\"{o}dinger Equation:
Singular Solutions and Optical Collapse} (Springer: Heidelberg, 2015).
\end{thebibliography}
\end{document}